\documentclass[fleqn,usenatbib]{mnras}

\usepackage{txfonts}

\usepackage[T1]{fontenc}
\usepackage{ae,aecompl}

\usepackage{graphicx}	
\usepackage{amssymb}	
\usepackage[utf8]{inputenc}

\title[{\tt StarHorse}: Stellar parameter estimation]{{\tt StarHorse}: A Bayesian tool for determining stellar masses, ages, distances, and extinctions for field stars}

\author[Queiroz et al.]{A. B. A. Queiroz$^{1,2}$\thanks{E-mail: anna.queiroz@ufrgs.br},
F. Anders$^{3,2}$,
B. X. Santiago$^{1,2}$,
C. Chiappini$^{3,2}$,
M. Steinmetz$^{3}$,
\newauthor
M. Dal Ponte$^{1,2}$,
K. G. Stassun$^{4}$,
L. N. da Costa$^{2,5}$,
M. A. G. Maia$^{2,5}$,
\newauthor
J. Crestani$^{1,2}$,
T. C. Beers$^{6}$,
J. G. Fern\'andez-Trincado$^{7,8}$,
D. A. Garc\'ia-Hern\'andez$^{9}$,
\newauthor
A. Roman-Lopes$^{10}$,
O. Zamora$^{9}$
\\\\
$^{1}$Instituto de F\'isica, Universidade Federal do Rio Grande do Sul, Caixa Postal 15051,Porto Alegre, RS - 91501-970, Brazil\\
$^{2}$Laborat\'orio Interinstitucional de e-Astronomia, - LIneA, Rua Gal. Jos\'e Cristino 77, Rio de Janeiro, RJ - 20921-400, Brazil\\
$^{3}$Leibniz-Institut f\"ur Astrophysik Potsdam (AIP), An der Sternwarte 16, 
14482 Potsdam, Germany\\
$^{4}$Vanderbilt University, Department of Physics \& Astronomy, VU Station B 1807, Nashville, TN 37235\\
$^{5}$Observat\'orio Nacional, Rua Gal. Jos\'e Cristino 77, Rio de Janeiro, RJ - 20921-400, Brazil\\
$^{6}$Department of Physics and JINA Center for the Evolution of the Elements, University of Notre Dame, Notre Dame, \\
IN 46556, USA\\
$^{7}$Departamento de Astronom\'ia, Universidad de Concepci\'on, Casilla 160-C, Concepci\'on, Chile\\
$^{8}$Institut Utinam, CNRS UMR6213, Univ. Bourgogne Franche-Comt\'e, OSU THETA,
Observatoire de Besan\c{c}on, BP 1615, \\
25010 Besan\c{c}on Cedex, France\\
$^{9}$Instituto de Astrof\'isica de Canarias, 38205 La Laguna, Tenerife, Spain\\
$^{10}$Departamento de F\'isica, Facultad de Ciencias, Universidad de La Serena, Cisternas 1200, La Serena, Chile
}

\date{Accepted XXX. Received YYY; in original form ZZZ}

\pubyear{2017}

\begin{document}
\label{firstpage}
\pagerange{\pageref{firstpage}--\pageref{lastpage}}
\maketitle

\begin{abstract}
Understanding the formation and evolution of our Galaxy requires accurate distances, ages and chemistry for large populations of field stars. Here we present several updates to our spectro-photometric distance code, that can now also be used to estimate ages, masses, and extinctions for individual stars. Given a set of measured spectro-photometric parameters, we calculate the posterior probability distribution over a given grid of stellar evolutionary models, using flexible Galactic stellar-population priors. The code (called {\tt StarHorse}) can acommodate different observational datasets, prior options, partially missing data, and the inclusion of parallax information into the estimated probabilities. We validate the code using a variety of simulated stars as well as real stars with parameters determined from asteroseismology, eclipsing binaries, and isochrone fits to star clusters. Our main goal in this validation process is to test the applicability of the code to field stars with known {\it Gaia}-like parallaxes. The typical internal precision (obtained from realistic simulations of an APOGEE+{\it Gaia}-like sample) are $\simeq 8\%$ in distance, $\simeq 20\%$ in age,$\simeq 6\%$ in mass, and $\simeq 0.04$ mag in $A_V$. The median external precision (derived from comparisons with earlier work for real stars) varies with the sample used, but lies in the range of $\simeq [0,2]\%$ for distances, $\simeq [12,31]\%$ for ages, $\simeq [4,12]\%$ for masses, and $\simeq 0.07$ mag for $A_V$. We provide {\tt StarHorse} distances and extinctions for the APOGEE DR14, RAVE DR5, GES DR3 and GALAH DR1 catalogues.
\end{abstract}




\section{Introduction}\label{introd}

Improving the accuracy and precision of stellar distances and ages, as well as individual interstellar extinction measurements, is one of the major tasks of stellar astrophysics in the {\it Gaia} era. Although the parallaxes from the first data release of the {\it Gaia} mission \citep{GaiaCollaboration2016}\footnote{\url{http://sci.esa.int/gaia/}} provide a major improvement for stars in the solar vicinity ($d\lesssim200$ pc), they do not yet reach the precision of spectro-photometric methods for the much larger distances probed by spectroscopic stellar surveys. Even after the final {\it Gaia} data release, foreseen for 2022, spectro-photometry will provide more precise distances for stars beyond 10 kpc. 

A large amount of spectroscopic 
data for individual stars has become available in recent years from dedicated surveys such as the Sloan Extension for Galactic Understanding and Exploration \citep[SEGUE,][]{Yanny2009}, the Apache Point Observatory Galactic Evolution Experiment \citep[APOGEE,][]{Majewski2017}, the RAdial Velocity Experiment (RAVE; \citealt{Steinmetz2006}), the Galactic Archaeology with HERMES survey (GALAH; \citealt{Martell2017}), the LAMOST Experiment for Galactic Understanding and Exploration \citep[LEGUE,][]{Deng2012}, and the Gaia ESO Survey (GES; \citealt{Gilmore2012}). The combination of such datasets with broad-band photometric data and the astrometric solutions from {\it Gaia} allow for a much more detailed modelling of the chemo-dynamical history of the Milky Way. On the one hand, {\it Gaia}'s proper motions and parallaxes, complemented with radial-velocity measurements, enable us to measure stellar phase-space distribution functions with unprecedented precision over a Galactic volume of $\sim 8000 {\rm kpc}^3$. {\it Gaia}'s parallaxes (in combination with spectroscopy) also help to estimate stellar masses and ages. And for more distant populations, more accurate spectro-photometric distances can be achieved by improved calibrations in the {\it Gaia} volume.
Such distance estimates are indispensable for mapping the chemical and kinematical properties of Galactic populations using large stellar surveys \citep[e.g.,][]{Boeche2013A,Boeche2014,Anders2014,Recio-Blanco2014,Nidever2014,Mikolaitis2014,Hayden2015,Carlin2015,Bovy2016c}. If stellar ages are available as well, they can be used to probe the chemo-dynamical evolution of different Galactic components much more directly \citep[e.g.,][]{Zoccali2003,Haywood2013,Mitschang2014,Anders2017,Mackereth2017}.

In \cite{Santiago2016}, we presented a Bayesian inference code to determine spectro-photometric distances for large survey samples, both in the optical and near-infrared (NIR). Since then, our group has extended that algorithm in several ways, improving the code's flexibility for different input data, updating priors and likelihood functions, and adding extinction, ages, and masses as parameters to be inferred by the method. In this paper we demonstrate these new capabilities. The paper is structured as follows: In Sec. \ref{method} we describe our method to estimate stellar parameters, distances, and extinctions. Section \ref{sec:updates} presents the recent updates to our code. We provide an analysis of the performance of our code in terms of internal accuracy and precision in Sec. \ref{simul}, using simulated stars, focussing especially on the new parameters mass, age, and extinction. We also discuss how biased spectroscopic parameters influence the estimated quantities. In Sec. \ref{validation} we compare our distances to several previous mass, age, and distance determinations that can be used as a reference. In Sec. \ref{dataproducts} we describe a {\tt StarHorse} application to a few spectroscopic surveys, with the purpose of delivering public releases of distances and extinction, We refrain from releasing ages and masses for the time being since their accuracy is still dependent on availability of Gaia-DR1 parallaxes and additional improvements. Gaia-DR2 will certainly improve their application to large volumes and datasets.  We conclude the paper with a summary and future plans in Sec. \ref{conclusion}.

\section {The Method}
\label{method}
\smallskip

Our method uses a set of spectroscopically-measured stellar parameters (typically effective temperature, $T_{\mathrm{eff}}$, surface gravity, $\log g$, and overall metallicity [M/H]), photometric magnitudes, $m_{\lambda}$, and parallax, $\pi$, to estimate the mass, $m_{\ast}$, age, $\tau$, distance, $d$, and extinction (in V band, $A_V$) for individual stars. The measured quantities are compared to predictions from stellar evolutionary models, following a statistical approach that is similar to previous works \citep[e.g.,][]{Burnett2010, Burnett2011, Binney2014a}, and that generalises the method presented in \citet{Santiago2016}.

The calculations rely on three important assumptions: Most importantly, we assume that the stellar models are correct, which might not be true for metal-poor stars as well as other limitations in the current stellar models, i.e., that the object of interest follows a canonical single-star evolutionary track. We caution that this assumption, even if a star is apparently single, can be violated to various degrees in practice, leading first and foremost to incorrect stellar mass and/or age estimates \citep[e.g.,][]{Brogaard2016, Yong2016, Fuhrmann2017a, Lagarde2017}. The second assumption is that the observational uncertainties of the measured parameters follow a normal distribution. The third assumption is that the observed measurements are independent, a condition that can also be violated in practice.

We can then calculate the probability that a set of independent measured parameters $\vec{x}=\{x_1, ..., x_n\}$ with associated Gaussian uncertainties ${\vec{\sigma_x}}$
is drawn from a set of theoretical values ${\vec{x_0}}$, predicted by some
model $\mathcal{M}$, by writing:
$$
P({\vec{x}},{\vec{\sigma_x}} \vert {\vec{x_0}})  = \prod_{i} \frac{1}{\sqrt{2\pi}\ \sigma_{x_i}} \exp \left[-\frac{(x_i-x_{0_i})^2}{2\sigma_{x_i}^2}\right].
\eqno{(1)}
$$
The above expression is called the likelihood of measuring the set $\{\vec{x}, \vec{\sigma_x}\}$ given a model $\mathcal{M}(\vec{x_0}[\vec{\theta}])$. Using Bayes's theorem, we now compute the posterior probability distribution (the probability of the model, given the data) as:

$$ P({\vec{x_0}} \vert {\vec{x}},{\vec{\sigma_x}}) = \frac{{P({\vec{x}},{\vec{\sigma_x}} \vert {\vec{x_0}})~P({\vec{x_0}}) }}{P(\vec{x},\vec{\sigma_x})}. \eqno (2)$$

\noindent The numerator contains the likelihood and the model priors $P({\vec{x_0}})$, and the denominator is the marginalized likelihood. It depends only on the measured parameters, being a constant through all the models that can be normalised out. 

In our case, the model family $\mathcal{M}(\vec{\theta})$ consists of a grid of stellar models computed for different ages, metallicities, and initial masses, convolved with a grid of distances and extinctions (modifying the apparent magnitudes of each stellar model). To evaluate the probability of some specific model quantity, $\vartheta$ (usually one that cannot be measured directly), we now compute the marginal posterior probability distribution function (PDF) for this quantity, by integrating over all variables of Equation (2), except $\vartheta$: 
$$
p(\vartheta) \coloneqq P(\vartheta|{\vec{x}},{\vec{\sigma_x}}) = \int \mathrm{d}x_{0, 0} \dots \mathrm{d}x_{0, n} P(\vec{x_0}[\vec{\theta}]\vert \vec{x},\vec{\sigma_x}). \eqno (3) 
$$

As mentioned earlier, a typical set of measured parameters includes $\vec{x} = 
\{\mathrm{[M/H]}, T_{\mathrm{eff}}, \log g, m_{\lambda}, \pi \}$, or any subset of these. The model parameters $\vartheta$ we compute are mass, $m_{\ast}$, age, $\tau$, distance, $d$, and $V$-band extinction, $A_V$. Our code delivers various statistics for the desired quantities. As in \cite{Rodrigues2014} and \citet{Santiago2016}, for each quantity we compute the median of the marginalised posterior probability distribution, $p(\vartheta)$ (Eq. 3), along with its 5\%, 16\%, 84\% and 95\% percentiles.

\section{Code Updates}\label{sec:updates}

In this section, we explain the technical details of our code in more detail (for an overview see the flow diagram in Fig. \ref{SHflux}). We encourage the reader to contact the developers\footnote{Anna Barbara Queiroz, Email: anna.queiroz@ufrgs.br \label{fot:email}} \footnote{Friedrich Anders, Email: fanders@aip.de\label{fot:email2}} for any questions or further details about the code. Via a parameter file, the user can choose the set of stellar models to be used, the available photometric and spectroscopic data, the treatment of extinction (whether to correct photometry for reddening or whether to include extinction as a parameter to be estimated), and the set of priors, among other options. Once the evolutionary models and the data are read in, the code operates according to the options chosen in the parameter file. These options, along with the other updates since \cite{Santiago2016}, are detailed in the next subsections. Readers interested only in the overall performance of the code may skip these.

\subsection{Including Parallax as a measured parameter}
\label{likepar}
To adapt our method to the new era of astrometric surveys like {\it Gaia}, JASMINE \citep{Gouda2012}, VLBA \citep{Melis2014}, and SKA \citep{Imai2016}, we introduced parallax as an optional measured input parameter for our code. As explained in the previous section, the likelihood can be extended for a generic group of measured parameters, so the method presents no difficulties to introduce the parallax in the likelihood, and it allows for much more precise estimates of stellar masses, ages, and extinction. When the user decides to use parallaxes as the primary input, we fix the range of distances for all models to be consistent with that measurement within $3\sigma$ (see \S \ref{likedist} below). If this is not specified, the possible range of distances to be probed for each stellar model is derived by matching an observed apparent magnitude, $m_{\lambda}$ (within $\pm 3 \sigma_{m \lambda}$), in some filter to the corresponding model absolute magnitude.

\subsection{Stellar parameters posterior}
Currently our code can determine distances, ages, masses, and extinctions, given a set of measured parameters by marginalising the joint posterior PDF. Below we explain in more detail how we build the values of distance, extinction, ages, and masses covered by the PDF.

\subsubsection{Distance}
\label{likedist}
If no reliable parallaxes are available, or if they are only available for a subset of stars, the range of distances to be probed comes from the available measured apparent magnitudes. We choose a master filter, $\lambda$, and create an array of length $N_d$ that ranges from $m_{\lambda 0} \pm 3\sigma_{m_{\lambda 0}}$, where $m_{\lambda 0}$ stands for intrinsic measured apparent magnitudes in the master filter. For each value of this array we compute the distance modulus, $(m_{\lambda 0}-M_{\lambda})$, for the absolute magnitudes in the model grid; these values are then finally transformed into an array of possible distances, $d$. When $A_V$ is not being estimated together with distance, the code assumes that the given magnitudes are previously corrected by the known extinction. See Section \ref{ext} for details of the estimation of $A_V$.
As explained in \S \ref{likepar}, if the user decides to use parallax measurements as the primary input, we build the distance array by inverting the array of allowed parallaxes. We then transform $d$ to intrinsic distance moduli that do not depend on colour or extinction. The $d$ and $(m-M)$ values are then used in the priors and likelihood to build the posterior PDF.

\subsubsection{Extinction}
\label{ext}
When multi-band photometry over a sufficient wavelength range is available, one can use the measured colours of a star to estimate interstellar dust extinction. When the intrinsic magnitudes are constrained by spectroscopic measurements, this extinction measurement can become very precise \citep[e.g.,][]{Rodrigues2014}. Our code can now also be used to determine extinction towards stars, by adding another free dimension to the model space.

When parallaxes are not available, we build a distance moduli that comes from the apparent magnitude $(m_{\lambda}-M_{\lambda})$, as explained in \S \ref{likedist}, but now the measured magnitudes are assumed not to be intrinsic. The distance moduli must then be corrected by an a priori unknown extinction: $(m_{\lambda 0}-M_{\lambda}) = (m_{\lambda}-M_{\lambda})-A_{\lambda}$. For each stellar model and each possible distance modulus, $(m_{\lambda}-M_{\lambda})$, we thus create $N_{A_V}$ random $A_V$ values from a previously defined range of possible $A_V$. If there is no initial guess of the $A_V$ for the given star, this range of $A_V$ values is kept fixed as $[-0.1,3.0]$. If some expectation for $A_V$ is available (an $A_V$ prior, $A_{V,0}$), we probe extinction in the range $[A_{V,0}/3,3\cdot A_{V,0}]$. We then transform the $A_V$ to $A_{\lambda}$ values using a chosen extinction curve (\citealt{Schlafly2016} by default; see Sect. \ref{schlafly} for a discussion, and subtract it from $(m_{\lambda}-M_{\lambda})$. Since the model space is large, we usually use $N_{A_V}=3$ to lower the computational cost. As long as the spectroscopic measurements do not confine the solution to a very small volume in model space, the marginalised PDFs over extinction and distance remain well-sampled.

When parallax information is available, the dereddened distance modulus array, $(m-M)_0$, is determined directly from the parallax. To determine the extinction we then use the reddened distance modulus arrays built from the apparent magnitudes, $(m_{\lambda}-M_{\lambda})$, and the difference between those two naturally delivers $A_{\lambda}$. 

\subsubsection{Masses and Ages}

Because masses and ages are quantities provided by the grid of evolutionary models, they are simply repeated over the additional dimensions of distance and extinction. Therefore, once we have a PDF from equation (3), we can directly estimate these parameters by marginalising over the distance and extinction dimensions.

\begin{figure*}
  \includegraphics[width=16.5cm]{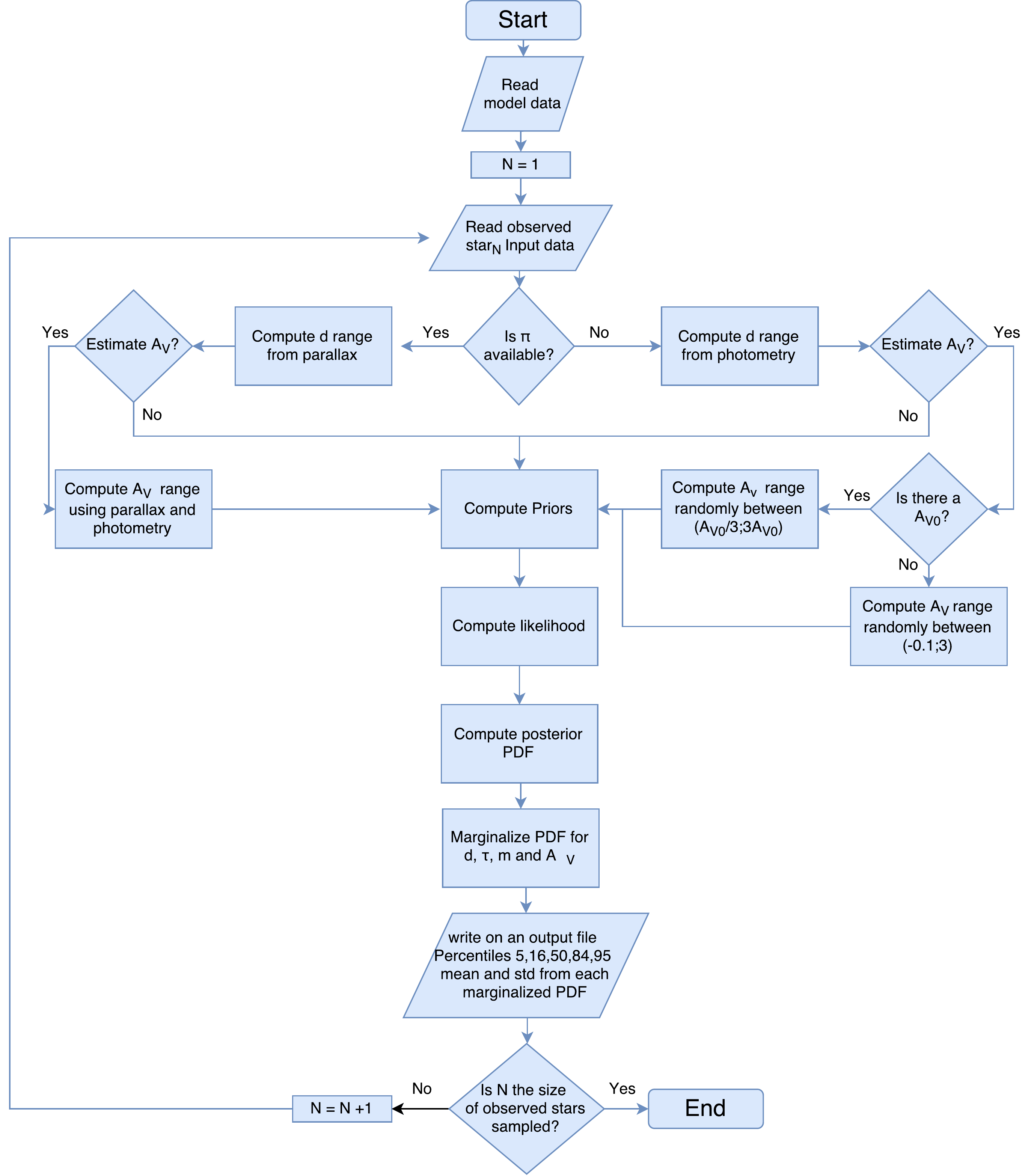}
     \caption{{\tt StarHorse} flux diagram. See Sec. \ref{method} and \ref{sec:updates} for details.}
     \label{SHflux}
\end{figure*}

\subsection{New spatial, chemical, and age priors}
\label{priors}

Our code uses several priors that summarise our prior knowledge about the initial mass function and the stellar population structure of the Milky Way, including the thin and the thick disk, and the stellar halo. For each Galactic component, these priors include the spatial density distribution, a metallicity distribution function (MDF) and an age distribution. In \cite{Santiago2016}, we adopted the same priors as \cite{Binney2014a}. In the current version, we updated our default structural parameters (solar position, scale lengths, scale heights, normalisations) from \citet{Binney2014a} to \citet{Bland-Hawthorn2016}.

Since APOGEE and other surveys are now probing also the inner regions of the Milky Way, we have also added simple spatial priors for the bulge/bar. The simplest choice of bulge spatial prior is a spherical exponential model with a exponential e-folding length of $0.5-1.0$ kpc. We also added the oblate model described by \cite{Dehnen1998}. Finally, as our default model we included a bar-like bulge model from \cite{Robin2012a}. Their models assume the bulge to be a triaxial ellipsoid, either boxy or disky (or yet a combination of both depending on the plane of projection), and with density laws that can be a $sech^2$, an exponential, or a Gaussian. Our code has been tested with one- and two-component model priors. Our default model is the S ellipsoid (bar component) taken from the ``S+E'' case listed in Table 2 of \cite{Robin2012a}. This ``S+E'' model was the minimum likelihood one among the models presented by the authors. The E component was removed based on the revision of the thick-disk structure and its extrapolation towards the inner Galactic regions made by \cite{Robin2014}, which effectively rendered the classical ellipsoid bulge unnecessary. We refer the interested reader to \cite{Robin2012a} and \cite{Robin2014} for more details.

\begin{figure}
  \resizebox{\hsize}{!}{\includegraphics[trim=0cm 2.5cm 0cm 0cm, clip=true]{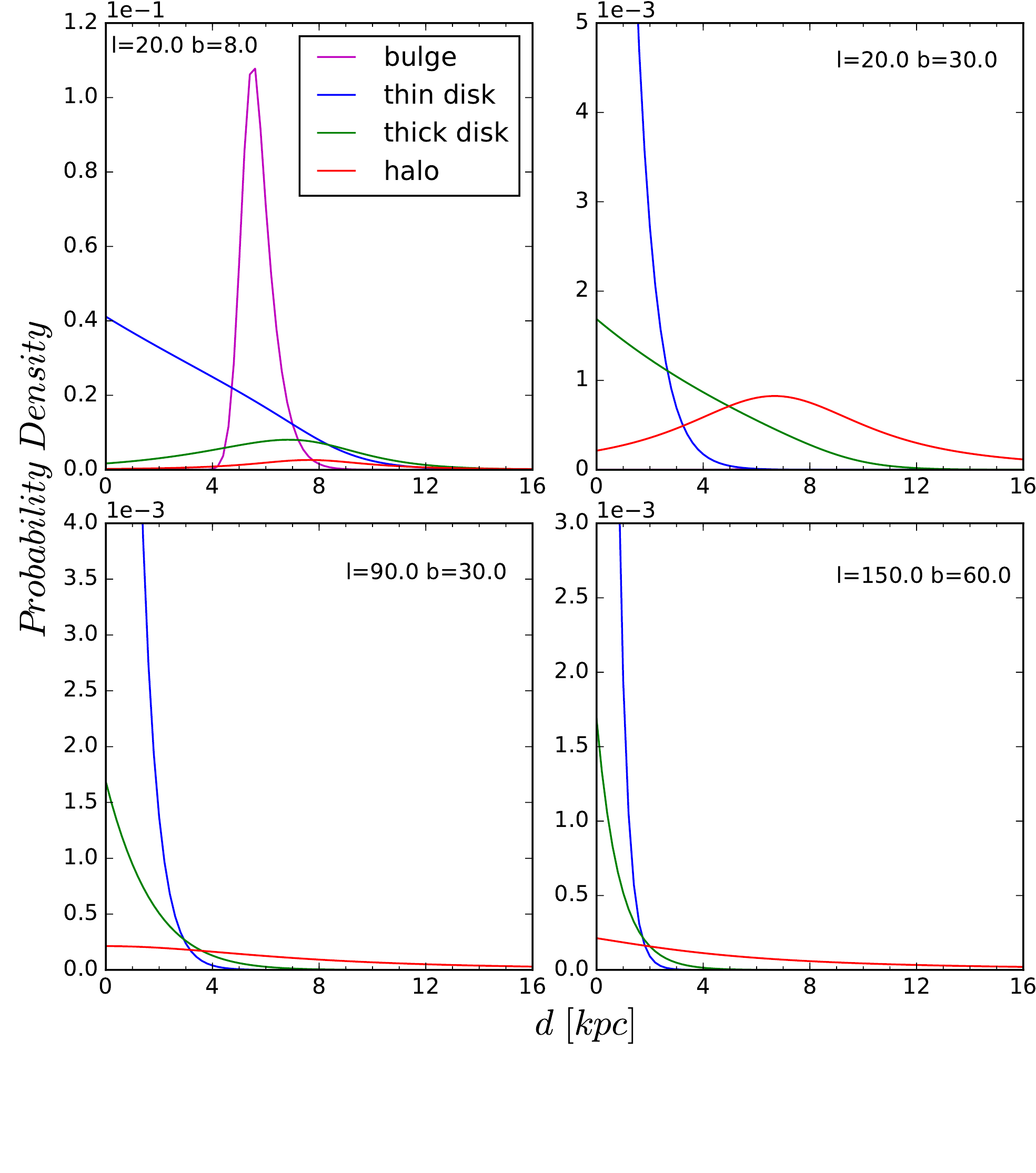}}
  \caption{Spatial profile for the Galactic components used in the {\tt StarHorse} priors. Each panel shows the behaviour of the prior probability density with distance for a given direction in Galactic coordinates ($l,b$).}
  \label{spatialprior}
\end{figure}

Fig. \ref{spatialprior} shows the contributions of each Galactic component spatial distribution as a function of distance, for four representative directions. 
In the upper left panel we show a direction towards the inner Galaxy $(l,b)=(20,8)$; in this case the bar/bulge component is the dominant population at heliocentric distances of $\sim 6$ kpc, its density decreases rapidly toward greater distances. The upper right panel shows a direction toward the same Galactic longitude, but at higher latitude. In this case the we miss the bulge, and notice that each of the other components dominates at a certain distance range.
The lower panels shows directions away from the Galactic centre. In the lower left panel $(l,b)=(90,30)$, we see that the contribution for the disks dominates out to 3.5 kpc, in the lower right panel $(l,b)=(150,60)$ the halo dominates already for $d>2$ kpc.

Our new age and metallicity priors for the four Galactic components are all assumed to be Gaussians. The corresponding mean and standard deviation values for each case are provided in Table \ref{priorgauss}. The motivation for this change is twofold: 1 - simplicity: they are simple functions, easily computed, which makes them ideal for the computationally intensive parameter estimate process used here; 2 - they are made broad enough to accommodate most or all of the recent age and metallicity distributions found in the literature, which are not only diverse, but also often conflicting. We assume that the impact of this choice is minor, though we are aware that the age and mettalicity distribution are not necessarly gaussians, by taking this approach, we avoid making our priors too specific, but do not completely overlook the knowledge accumulated about the different Galactic components. 

We note that the previous age priors from \cite{Santiago2016} assigned zero probability to disk stars older than 10 Gyrs and to thick disk or halo stars younger than this value. Recent results found in the literature pose a challenge to such simple age step functions. One example is the discovery of young $\alpha$-enhanced stars, likely thick-disk members \citep{Chiappini2015a}. The previous metalicity priors were also narrower, specially for the thin-disk, essentially ruling out any thin disk star more metal-poor than $[Fe/H] \simeq -0.6$. The changes made to these components also make them more in sync with our current understanding of the bulge populations, for which there is also recent evidence for a larger fraction of stars younger than $\simeq 5$ Gyrs \citep{Bensby2013a, Valle2015a}.

\begin{table}
\centering
\caption{Adopted parameters of the Gaussian age and metallicity priors for the Galactic components. }
\footnotesize
\setlength{\tabcolsep}{2pt}
\begin{tabular}{ccccc}
\hline
\hline
Component & Mean age & $\sigma$ age & Mean [M/H] & $\sigma$[M/H] \\
\hline 
Thin disk & 5.0 Gyr & 4.0 Gyr & -0.2  & 0.3 dex\\
Thick disk & 10.5 Gyr & 2.0 Gyr & -0.6  & 0.5 dex  \\
Halo & 12.5 Gyr & 1.0 Gyr & -1.6 & 0.5 dex \\
Bulge & 10.0 Gyr & 3.0 Gyr & 0.0 & 0.5 dex \\
\hline
\end{tabular}
\label{priorgauss}
\end{table}

\subsection{A new default extinction curve}\label{schlafly}

The adopted extinction curve, i.e. the dependence of the absolute extinction on wavelength, has of course an influence on the distances and extinctions provided by our code. 

To quantify this effect, we recently implemented the one-parameter extinction curve presented by \citet{Schlafly2016}, derived from APOGEE spectroscopy in combination with Pan-STARRS1, 2MASS, and WISE data, as an alternative to the \citet{Cardelli1989} curve that was determined based on very few stars. For the mostly low-latitude APOGEE DR14 sample, where the sensitivity of our results to the chosen extinction law is arguably highest, we tested the effect of changing the extinction curve from \citet{Cardelli1989} to \citet{Schlafly2016}. 

Our tests showed that, if both optical (APASS) and near-infrared magnitudes are available (i.e. for the brighter stars), the systematic effect of the adopted extinction curve is almost negligible: median differences in the inferred parameters range in the $1\%$-percent regime, although there are some weak systematic trends with age. The main differences in the inferred parameters arise when only 2MASS magnitudes are available: in this case the results using \citet{Schlafly2016} yield $\sim5\%$ lower $A_V$, $\sim9\%$ greater ages, $\sim0.6\%$ greater distances, and $\sim11\%$ higher $A_V$. Since the \citet{Schlafly2016} extinction curve is much more realistic, and its adoption also slightly improves the convergence of the code for high-extinction stars (i.e. distances to $\sim 10,000$ more APOGEE DR14 stars could found), the default {\tt StarHorse} extinction curve is now \citet{Schlafly2016}. \\

\section{Tests with simulated stars}
\label{simul}
To validate our code, we carried out several tests using simulated stars. Since one of our main goals is to apply {\tt StarHorse} to field stars with reliable parallaxes to infer masses and ages, we include parallax in the set of observed quantities of our simulated stars, and use them to constrain our distance range.

The first sample of simulated stars we used is identical with the set used in \citet{Santiago2016}, except for the distance range and the assumed spectroscopic uncertainties. It consists of 5000 randomly drawn PARSEC models \citep{Bressan2012}, convolved with Gaussian errors in the spectro-photometric parameters and parallaxes (see details in Table \ref{sampletable}). The error values for the ``high-res'' version of this simulation were inspired by the spectroscopic uncertainties from the APOGEE DR14 results, while the low-res case was based on typical uncertainties from the SEGUE and RAVE surveys. In both cases the samples contain stars with distances between 0.05 and 1 kpc (typical {\it Gaia}-TGAS distances), random galactic positons, and extinctions from \cite{Schlegel1998}.

The PARSEC simulated stars are useful to map the internal accuracy and precision of our estimated distances, ages, masses, and extinctions over a wide input parameter range. The caveat is that the PARSEC sample sets are not representative of a real magnitude-limited sample of stars in our Galaxy. By randomly picking PARSEC models, we tend to oversample young stars, regardless of metallicity, relative to most survey data in the nearby Galaxy. This also means that our prior knowledge about Galactic stellar populations does not apply to these simulations, so that in this case we set all priors to unity.

\begin{table*}
\centering
\caption{Adopted geometry, local or central calibration and SFR+AMR for the Galactic components simulated with TRILEGAL}
\footnotesize
\setlength{\tabcolsep}{2pt}
\begin{tabular}{cccc}
\hline
\hline
Component & Spatial distribution & Local/Central Calibration & SFR+AMR \\
\hline 
Thin disk & Squared hyperbolic secant   & Local 55.40 $M_\odot pc^{-2}$ & 2-step age + \cite{Fuhrmann1998} + \\
& \cite{Girardi2005}  &  & $\alpha$-enh \cite{Girardi2005}\\
\hline
Thick disk & Squared hyperbolic secant & Local 0.001 $M_\odot pc^{-3}$ & 11-12 Gyr const. + \\
& \cite{Girardi2005} &  & Z = 0.008 with $\delta [M/H]$ = 0.1dex \\
\hline
Halo &  Power law & Local 0.0001 $M_\odot pc^{-3}$ &12-13 Gyr + \\
&  \cite{deJong2009} &  &\cite{Ryan1996} $[M/H]$ distribution \\

\hline
Bulge & Triaxial bulge  & Central 406 $M_\odot pc^{-3}$ & 10 Gyr +   \\
 & \cite{Vanhollebeke2009} &  &  \cite{Zoccali2003} $[M/H]$ + 0.3 dex \\
\hline
\hline
\end{tabular}
\label{tabletri}
\end{table*}

To test the code in more realistic scenario, we also use a TRILEGAL \citep{Girardi2012} population-synthesis simulation of an APOGEE-TGAS-like sample of giant stars. The details of this simulation are given in Table \ref{tabletri}; we describe the main features briefly here. The underlying stellar models of TRILEGAL are from \cite{Marigo2008}, which are similar but not identical to our default PARSEC 1.2S models. We used a \cite{Chabrier2003} log-normal initial mass function (IMF) for all Galactic components (thin disk, thick disk, bulge, and stellar halo), and the default spatial distribution, density normalization, star-formation rate (SFR), and age-metallicity relation (AMR) for all components (see Table \ref{tabletri}). Extinction was assumed to result from an exponential dust disk with calibration at infinity of $A_V=0.0378$ mag for the Galactic poles, and the photometry is in the $UBVRIJHK$ system \citep{MaizApellaniz2006}.
The Solar position and Solar height above the disk were assumed to be $R_{\odot}=8.7$ kpc and $Z_{\odot}=24.2$ pc, respectively  (deviant from the values of our Galactic priors, $R_{\odot}=8.2$ kpc and $Z_{\odot}=11.1$ pc; see Sec. \ref{priors}). To simulate APOGEE-TGAS-like observations, we convolved the TRILEGAL stellar parameters with Gaussian errors, as in the PARSEC ({\it high-res}) sample, and introduced cuts in $\log g$ and distance. The uncertainty values and stellar parameter ranges for TRILEGAL sample are again listed in Table \ref{sampletable}. 

The main aim of using this TRILEGAL simulation was to test the impact of the different Galactic priors in our parameter estimates. Therefore, we ran {\tt StarHorse} on the TRILEGAL sample for three prior configurations: i) no spatial, age, and metallicity priors, only the IMF; ii) IMF and spatial priors only; iii) all priors.

\begin{table*}
\centering
\caption{Summary of the reference data: parameter ranges, uncertainties, and provenance. }
\setlength{\tabcolsep}{3pt}
\begin{tabular}{cccccccccccc}
\hline
\hline
Sample & $\sigma(\pi)$  & $d$ range & $\sigma(T_{\mathrm{eff}})$ & $T_{\mathrm{eff}}$ & $\sigma \log g$ & $\log g$ & $\sigma$ [M/H] & [M/H]  & $\sigma$ mag & mag range & filters \\

 & [mas] &  [kpc]  & [K] &  range [K] & & range &  & range &  &  [V mag] & \\

\hline 
PARSEC high-res & 0.3 & $0.05-1$  & $70.0$ & $3000-7000$ & $0.08$ & $1-5$ & $0.03$ & $-2.5 - 0.5$ & 0.025 & $-2-24$ & $BVgriJHK_s$\\
PARSEC low-res & 0.3 & $0.05-1$  & $95.0$ & $3000-7000$ & $0.24$ & $1-5$ & $0.12$ & $-2.5 - 0.5$ & 0.025  & $-2-24$ & $BVgriJHK_s$ \\
TRILEGAL & 0.3 &$0.05-1$  & $70.0$ & $3000-7000$ & $0.08$  & $1-4.1$ & $0.03$ & $-2.5 - 0.5$ & 0.025 & $4 -13$  & $BVRIJHK_s$ \\
Detached Eclipsing Binaries & 0.04  &$0.01-65$  & $80$ & $4320-5730$  & $0.02$ & $1-3.6$  & $0.12$ & $-1.1 - 0.1$  & $0.03$ &  $0.7- 18$  & $BVRI$\\
Other Eclipsing Binaries & 0.36 & $0.03-1$ & $256.0$ & $3880-30300$ & $0.02$ & $2.9-4.5$ & $0.1$ & $0.03-0.2$ & $0.06$ & $5-12$  & $BVRIJHK_s$\\
CoRoGEE stars & 0.007 &$0.8-10$  & 90.0  & $4000-5500$ & 0.05 & $1.4-3.0$ & 0.03 & $-2.5 - 0.5$ & $0.025$ & $11-16$ & $BVgriJHK_s$\\
OCCASO clusters stars & 0.06 & $0.05-6$  & 60.0 & $4300-5300$ & 0.1 & $1.7-3.2$ & 0.2 & $ 0.0 - 0.4$ & $0.03$ & $8 - 15$ & $JHK_s$\\
\hline
\hline
\end{tabular}
\label{sampletable}
\end{table*}

\subsection{Internal accuracy and precision}

\begin{figure*}
  \centering
  \includegraphics[width=15.5cm, trim=1.0cm 2.3cm 2cm 0cm, clip=true]{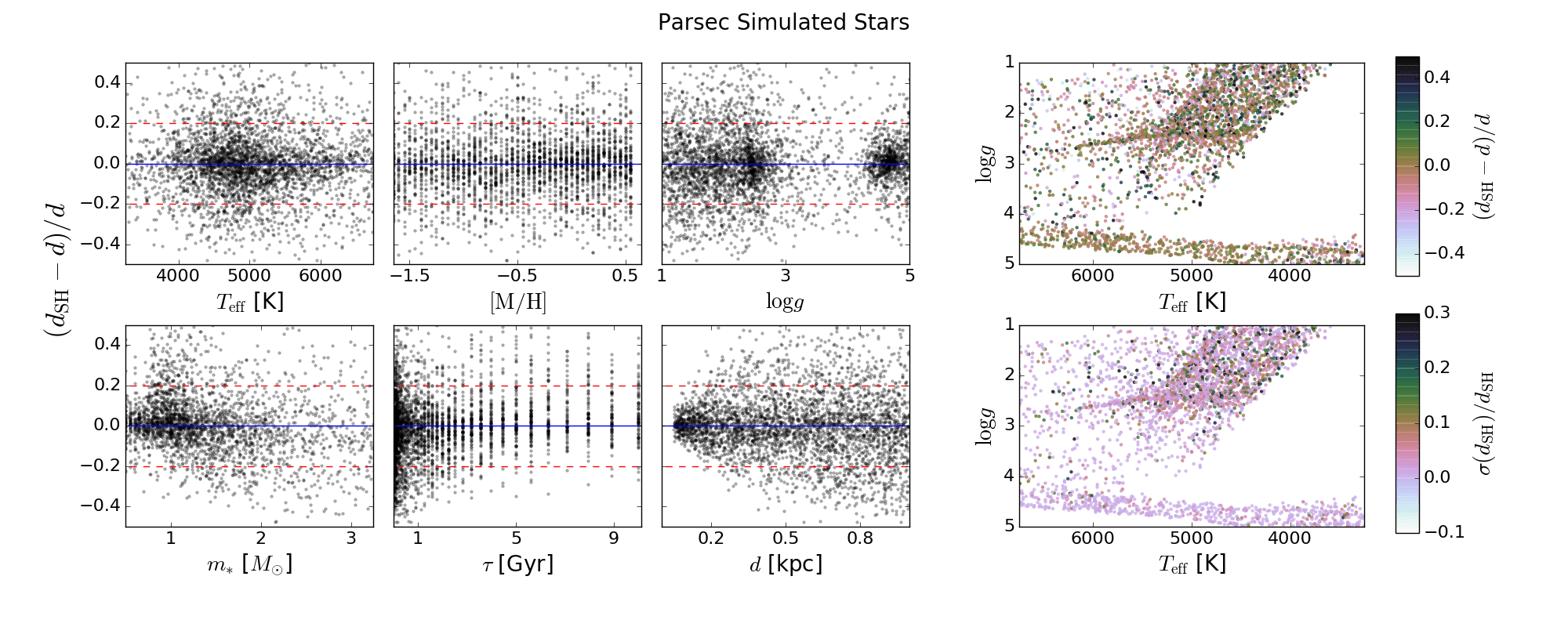}
  \includegraphics[width=15.5cm, trim=1.0cm 2.3cm 2cm 1.65cm, clip=true]{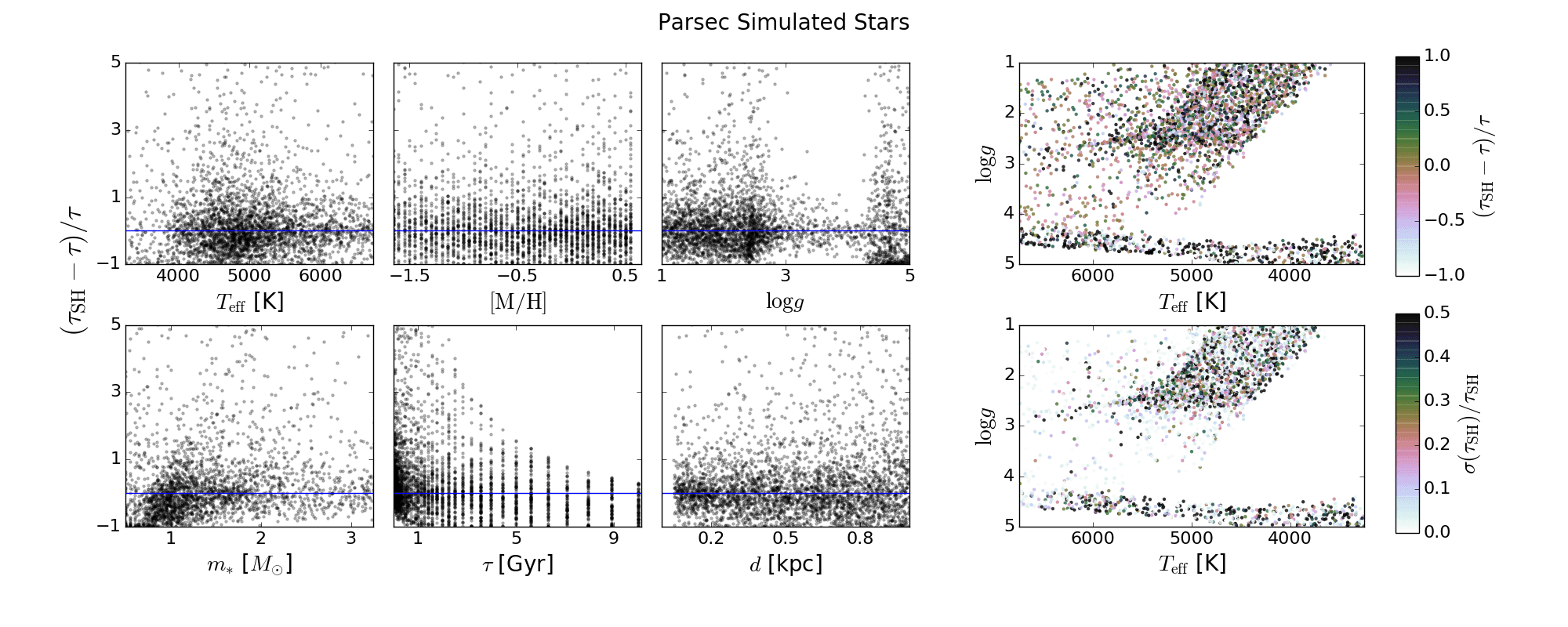}
  \includegraphics[width=15.5cm, trim=1.0cm 2.3cm 2cm 1.65cm, clip=true]{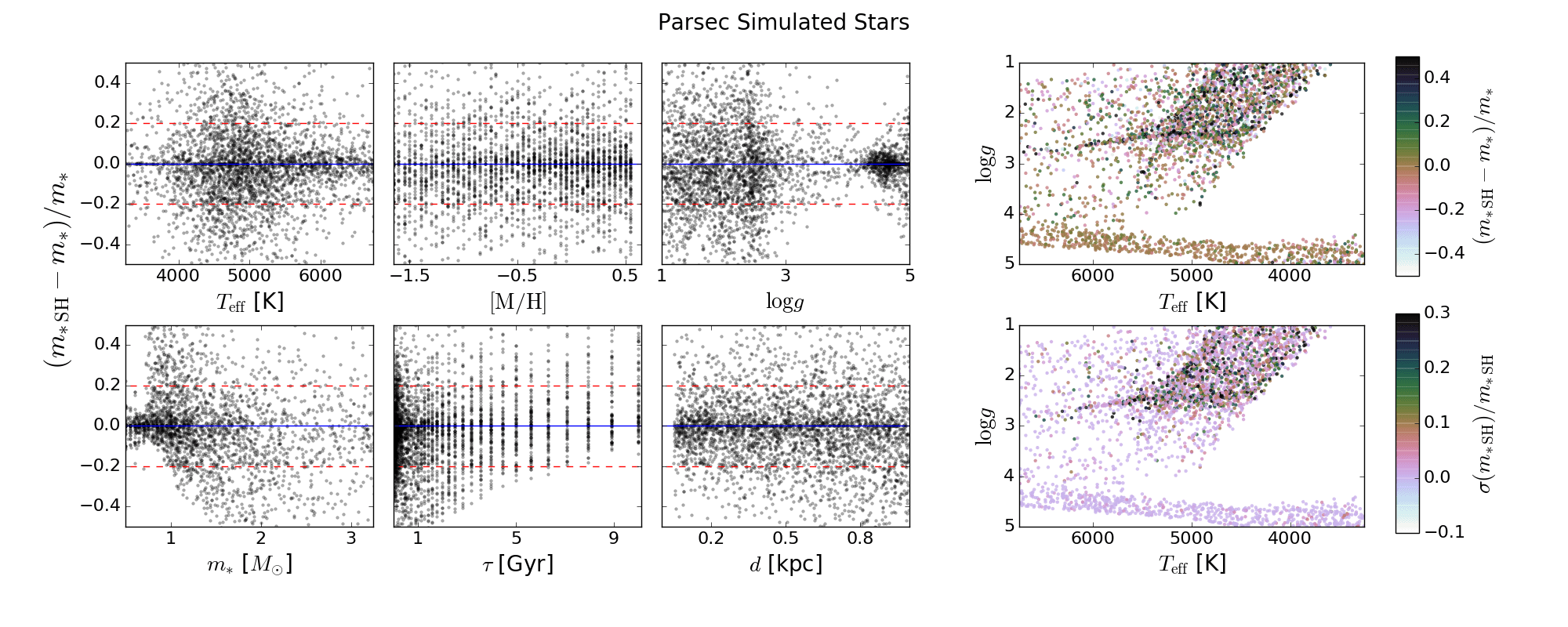}
  \includegraphics[width=15.5cm, trim=1.0cm 2.3cm 2cm 1.65cm, clip=true]{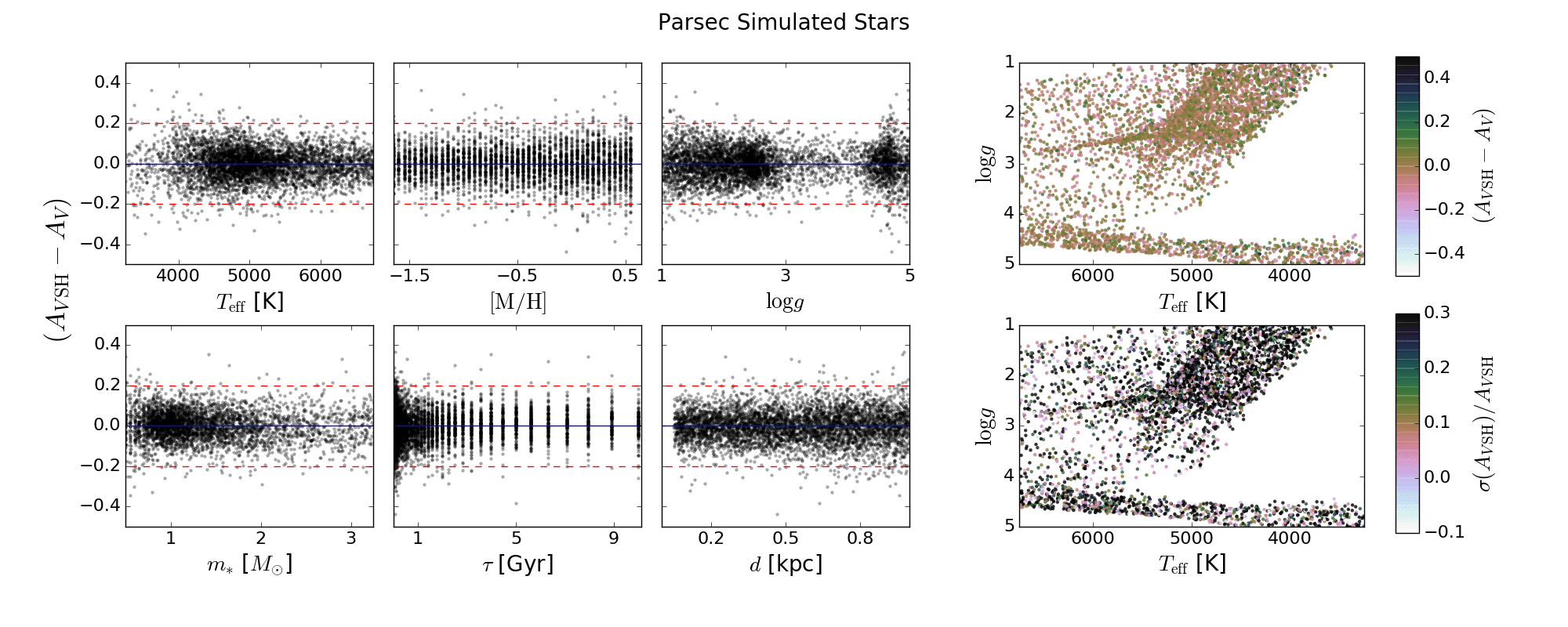}
  \caption{{\tt StarHorse} relative  errors for the PARSEC high-res simulations. For each parameter, the six panels on the left show the relative errors, $d$ (top rows), $\tau$ (third and fourth row), $m_\ast$ (fifth and sixth row), $A_V$ (last rows) as a function of the true parameters. The solid blue line is the identity line, and the dashed red lines correspond to $\pm 20\%$ errors (except for ages: 40\%). The panels on the right show the relative errors (top panel) and uncertainties (bottom panel) in the $\log g$~vs.~$T_{\mathrm{eff}}$ plane. } 
  \label{parsec}
\end{figure*}

Figure \ref{parsec} shows the results of our PARSEC simulated-star tests (high-res case). The first two rows show the relative errors in distance, $(d_{SH}-d_{True})/d_{True}$, where $d_{SH}$ are the distances estimated by our code (SH standing for {\tt StarHorse}). Each panel shows these same errors as a function of a different parameter. The last panels on the right show the relative distance errors (top row) and uncertainties (2nd row) mapped onto the $\log g~vs.~T_{\mathrm{eff}}$ diagram. The first line of Table \ref{perctable} shows the relative distance error values that correspond to the 5\% -ile, 16\% -ile, 50\% -ile, 84\% -ile, and 95\%-ile positions of the relative error distribution, when their signs are omitted. We also list the median value of the full error distribution, to quantify the presence of systematic trends. For example, we see that 50\% of the simulated stars have distance errors of less than 6\%, and 84\% have distance errors below 16\%. There is no strong systematic trend with any of the parameters, apart from an increase in the errors for larger distances and for lower $\log g$ values (giants). We also note that the discreteness of the PARSEC model grid used is visible in most of the panels.

The remaining rows of Fig. \ref{parsec} shows the same type of plots, but now with the relative errors in age, mass, and $A_V$. As before, the percentiles of the relative error distributions are listed in Table \ref{perctable}. As in the case of distances, the mass estimates (rows 5 and 6 in Figure \ref{parsec}) do not suffer from any clear systematics with the main parameters, apart from the trend of increasing errors with distance. There is a subset of mostly subgiant and dwarf stars ($\log g > 3.5$, and $m_{ast} < 0.8 m_{\odot}$) with very-well determined masses. From Table \ref{perctable}, we see that 50\% (84\%) of the estimated masses agree with the true values within 7 (22)\%, and that the outliers are predominantly young (massive) evolved stars, which are rare in the Milky Way field.

As for ages, shown in the 3rd and 4th rows of Figure \ref{parsec}, catastrophic errors, of 100\% or more, occur for about 15\% of the stars. These stars are not restricted to a small subset of parameter space. But age is the main parameter leading to these catastrophic errors, which are more frequent for $\tau < 1$ Gyr. There is also some dependence on distance (i.e., low parallaxes, for which the parallax error is relatively large) and mass (or $\log g$). Since dwarfs change their position in the spectroscopic Hertzsprung-Russell diagram only slightly on long timescales, ages for dwarfs are more uncertain; the age PDF will tend to be very flat.
The $A_V$ errors shown in the last two rows are within $0.05$ ($0.1$) mag for 50\% (84\%) of the stars, with no clear systematic effects.

\begin{figure*}
  \centering
  \includegraphics[width=15.6cm, trim=1cm 2.3cm 2.0cm 0cm, clip=true]{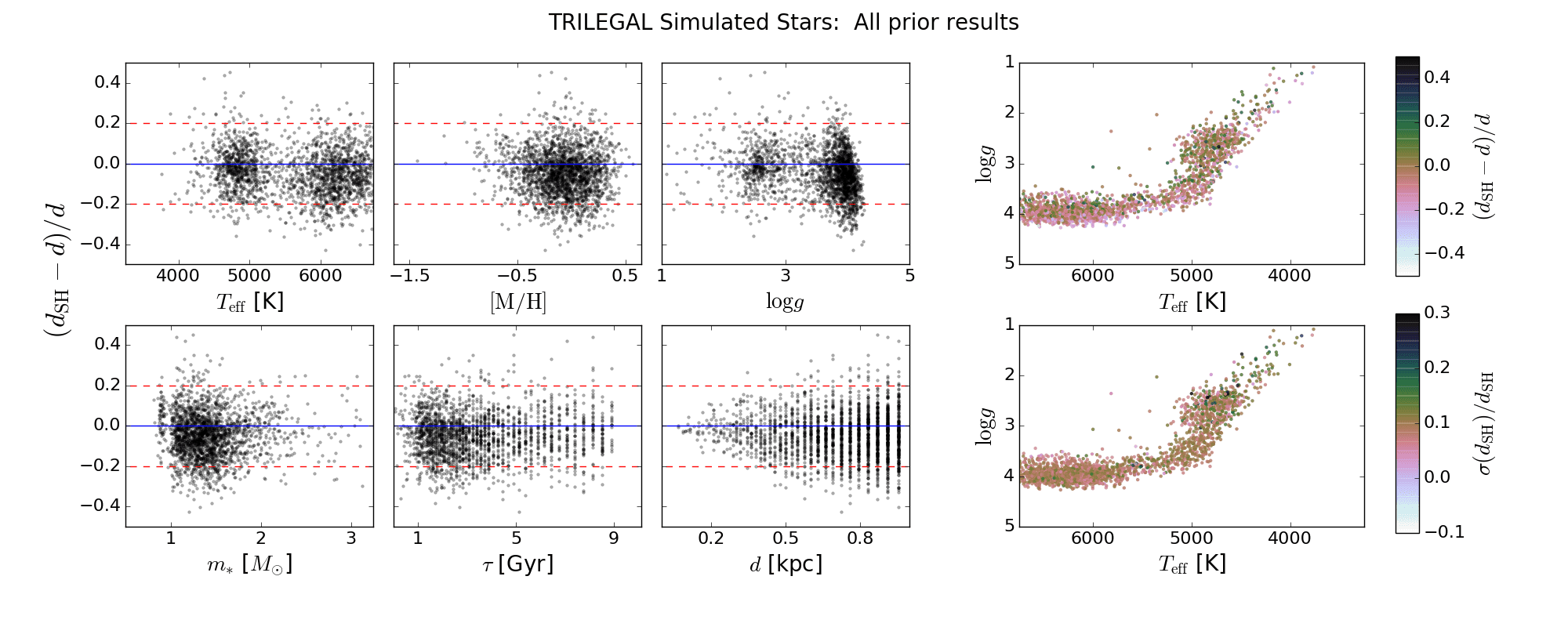}
  \includegraphics[width=15.6cm, trim=1cm 2.3cm 2.0cm 1.6cm, clip=true]{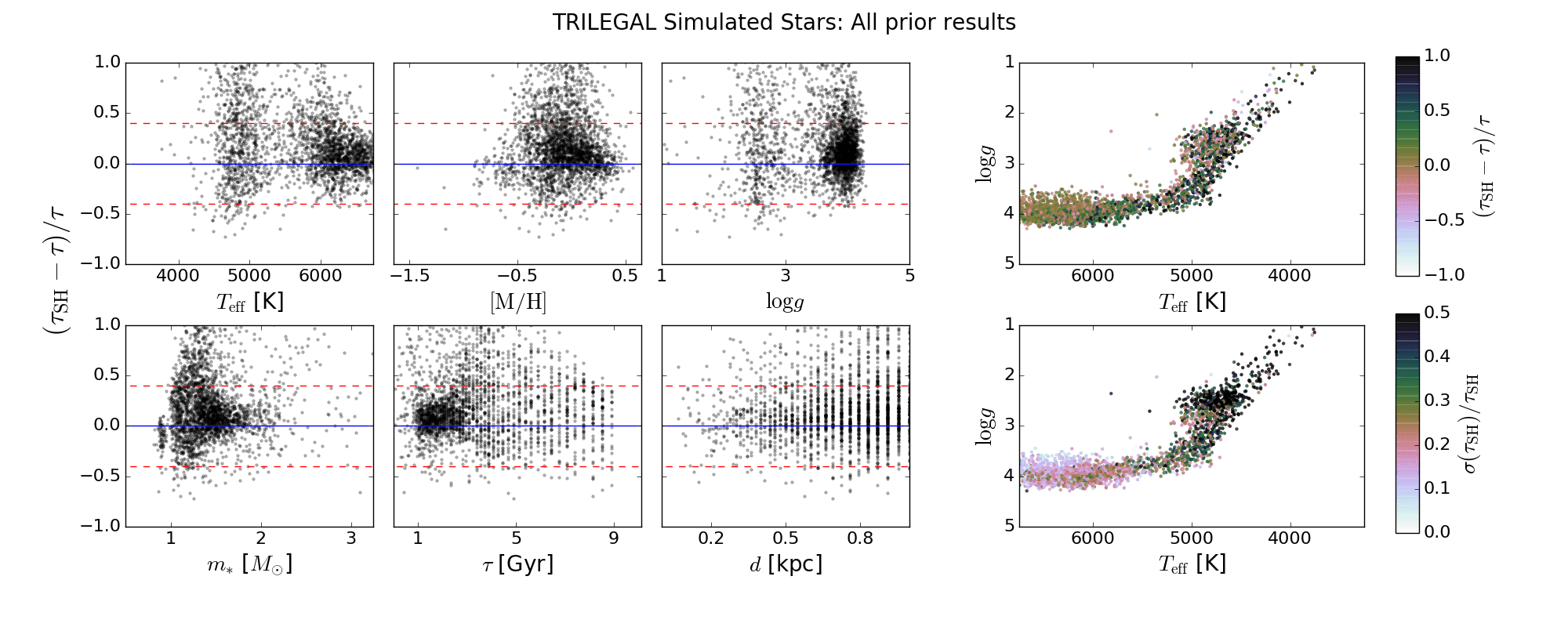}
  \includegraphics[width=15.6cm, trim=1cm 2.3cm 2.0cm 1.6cm, clip=true]{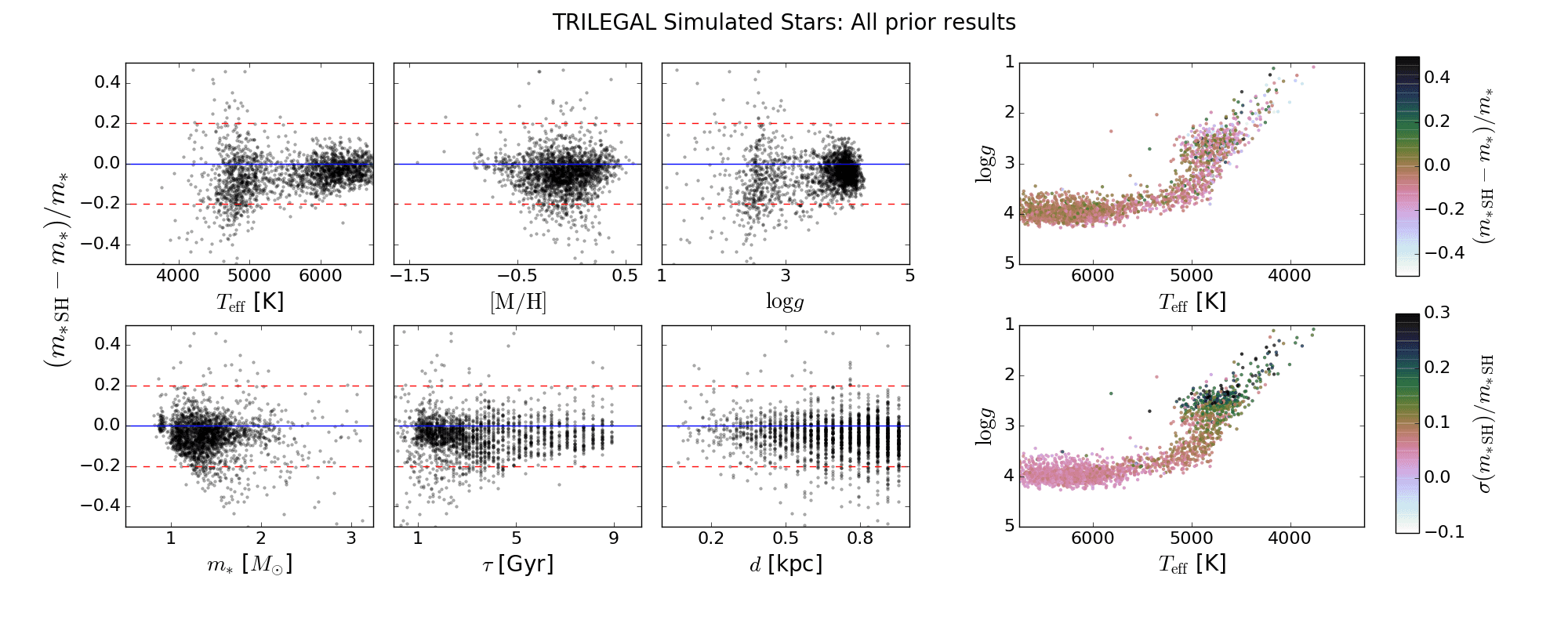}
  \includegraphics[width=15.6cm, trim=1cm 2.3cm 2.0cm 1.6cm, clip=true]{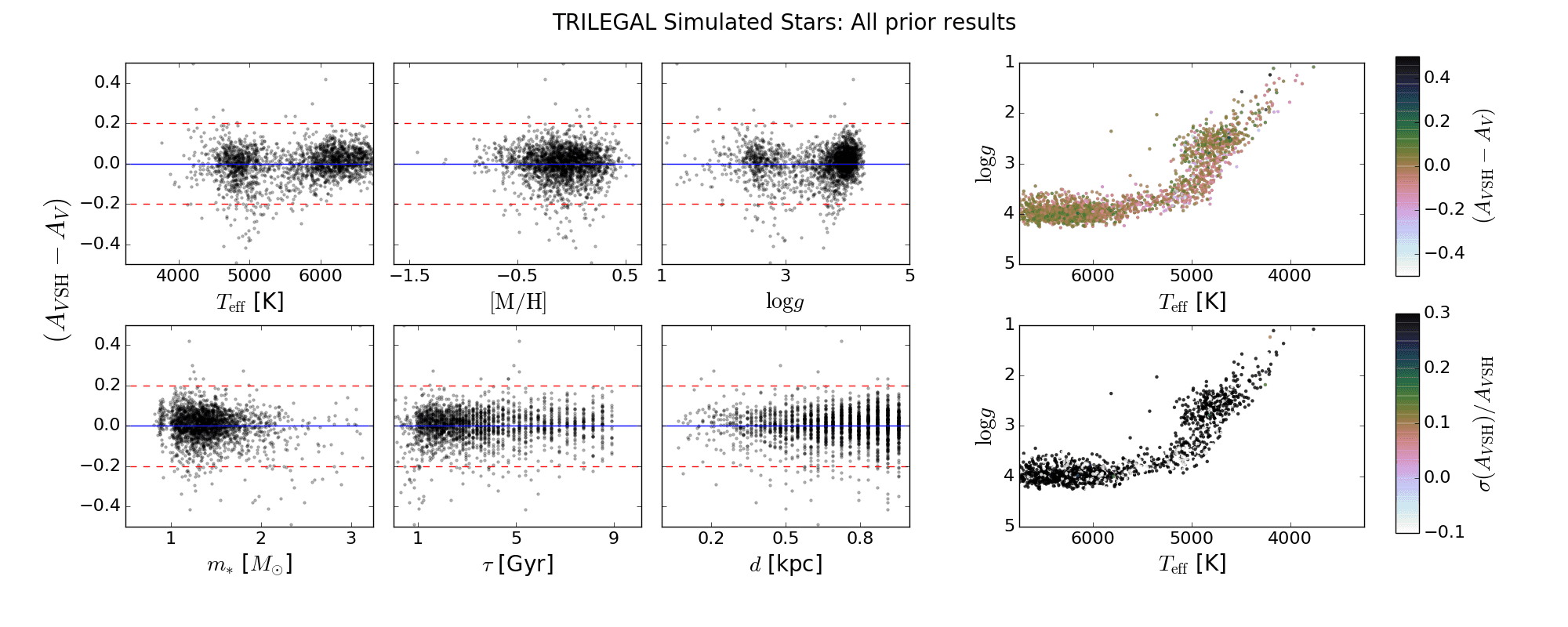}
  \caption{Relative distance, age, mass, and $V$-band extinction errors for the TRILEGAL simulations, in the same style as Fig. \ref{parsec}.} 
  \label{trilegalfieldbulge}
\end{figure*}

Figure \ref{trilegalfieldbulge} shows the results of our TRILEGAL test sample, in the same style as Fig. \ref{parsec}, for the case where all priors (IMF, spatial, age, and metallicity) are used. Although the stellar-parameter range of the TRILEGAL sample is similar to the PARSEC simulations, the resulting error distributions vary considerably when compared to the PARSEC results. There is a relative scarcity of low-mass and young stars in the TRILEGAL simulations when compared to PARSEC. Most dwarfs are removed by the APOGEE colour cut, and most red M dwarfs ($m_{\ast} < 0.8 m_{\odot}$, $\log g > 4$) are too faint to be seen. Likewise, most stars, even in the thin-disk component, are older than $\tau \simeq 500$ Myr. Relatively fast evolutionary stages, like post-HB phases, are also less frequent in the TRILEGAL simulations. 

The first two rows of Figure \ref{trilegalfieldbulge} show that there is a small trend towards underestimating distances overall, especially for hot dwarfs and subgiants (high $T_{\mathrm{eff}}$ and $\log g$ clouds in the upper row). Errors in all four parameters also tend to systematically increase with distance. Slight general trends are also observed in the sense of overestimating ages, and underestimating masses and extinction (these latter two parameters seem to be more affected for low $T_{\mathrm{eff}}$ stars). Still, typical errors are of order $\simeq 8\%$, $\simeq 19\%$, $\simeq 6\%$, and $\simeq 0.04$ mag, respectively, for distances, ages, masses and $A_V$ (Table \ref{perctable}). We also notice that the absence of a large number of young main sequence stars significantly reduces the occurrence of catastrophic age errors in TRILEGAL. 

\break

To test the influence of our prior knowledge, the TRILEGAL results shown in Fig. \ref{trilegalfieldbulge} (all priors included) can be compared to Fig. \ref{trilegalnoprior} in Appendix A. That figure shows the corresponding results for the case when no priors were adopted. In general, the results in the {\it all priors} and in the {\it no priors} cases are quite similar, indicating that the accuracy of our code is robust to prior assumptions. We also tested the case where only the spatial priors were used, but not the age and MDF ones. As expected for this intermediate case, the panels are again very similar to those shown in Figures \ref{trilegalfieldbulge} and \ref{trilegalnoprior}.

\begin{figure*}
  \centering
  \includegraphics[width=17cm]{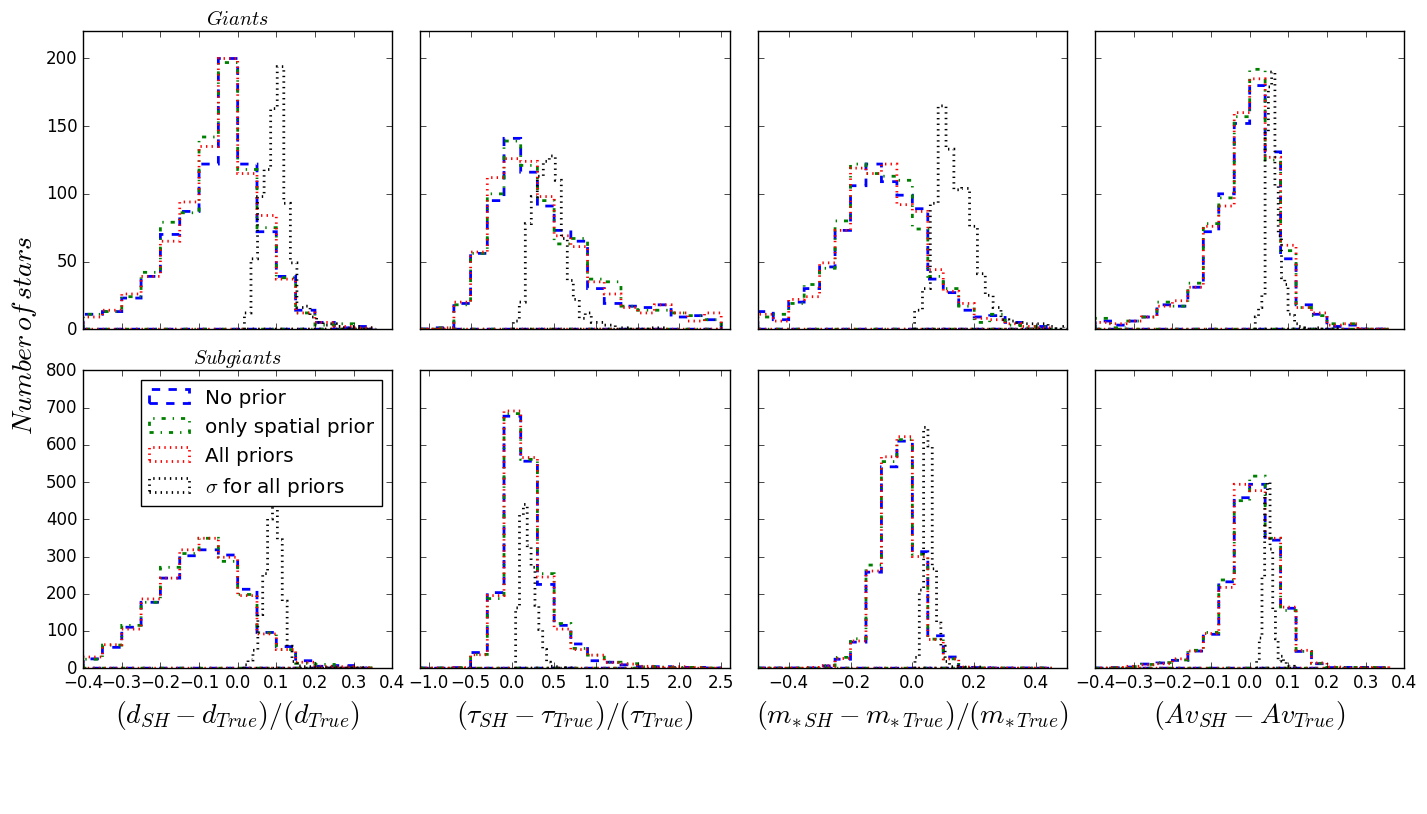}
  \caption{Testing priors with TRILEGAL simulations. The figure shows the distributions of relative errors and uncertainties in distance (leftmost column), age (second column from left), mass (third column from left) and $A_V$ (rightmost column). The top panels show the results for giant stars ($\log{g}<3.5$) and the bottom ones are for subgiant stars ($4.1>\log{g}>3.5$). The different lines in each panel correspond to the 3 different prior assumptions used. The blue line represents the results when no spatial, age and metallicity priors are adopted, the green line is based on spatial priors only, and the red line shows the results for all priors. The black lines represent the distribution of {\tt StarHorse} uncertainties normalised by the parameter, only for the all priors case.} 
  \label{figtri}
\end{figure*}

This is further corroborated by Figure \ref{figtri}, where we show the relative error distributions of age, distance, mass, and $A_V$ for the three combinations of priors, and separated for subgiants/hot dwarfs and giants. The histograms confirm the trends seen in the scatter plots, and they also show that the parameter estimates are not strongly dependent on the priors adopted, at least out to our maximum distance of 1 kpc (TGAS volume).
We note that the spatial density profiles, SFH and MDF used by TRILEGAL are not the same as those in {\tt StarHorse}, which shows the importance of adding basic and non-restrictive priors in the parameter inference.


\subsection{Effect of systematic stellar parameter errors}\label{sys}

\begin{figure*}
  \centering
  \includegraphics[width=13.8cm, trim=0cm 0cm 0cm 0cm, clip=true]{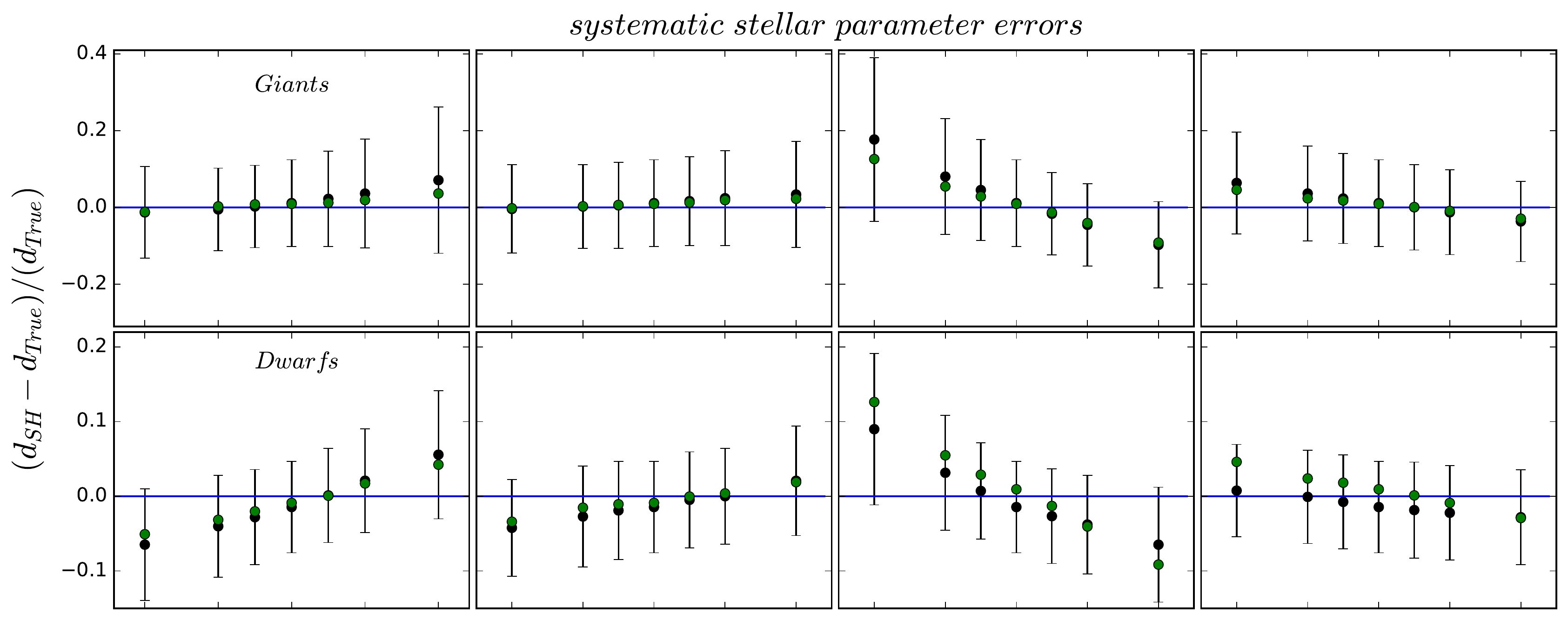}
  \includegraphics[width=13.8cm, trim=0cm 0cm 0cm 0cm, clip=true]{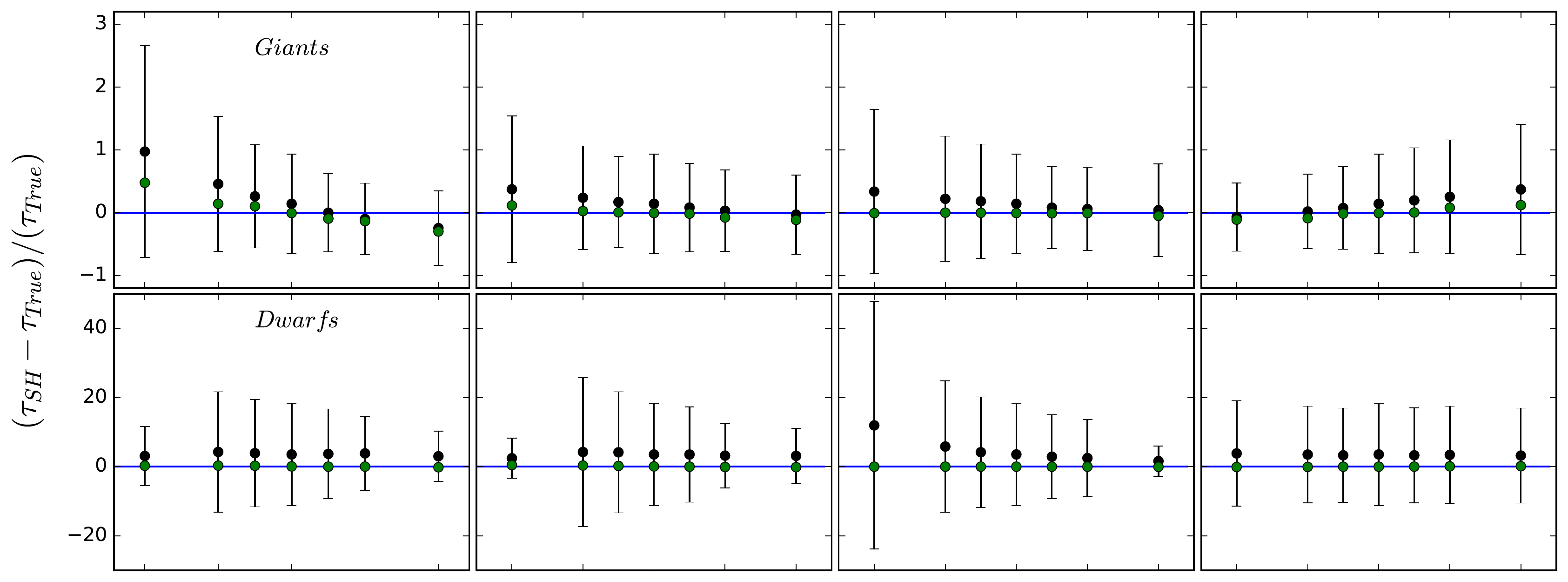}
  \includegraphics[width=13.8cm, trim=0cm 0cm 0cm 0cm, clip=true]{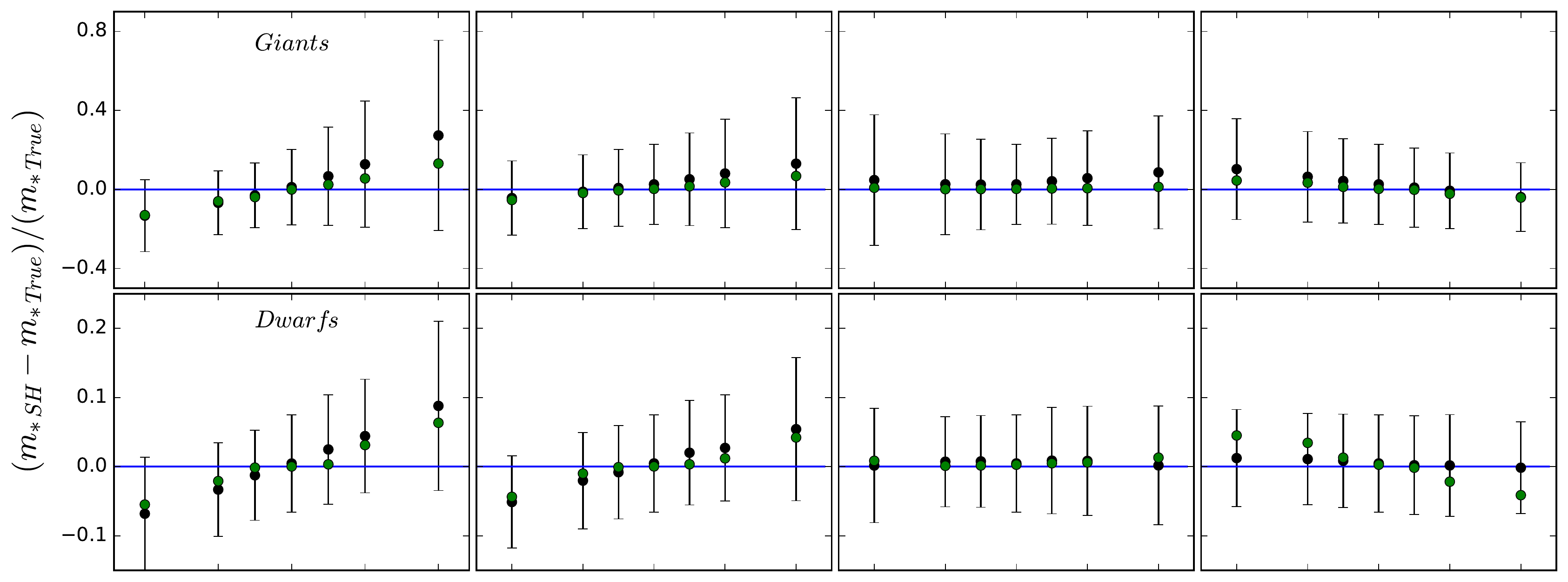}
  \includegraphics[width=13.8cm, trim=0cm 0cm 0cm 0cm, clip=true]{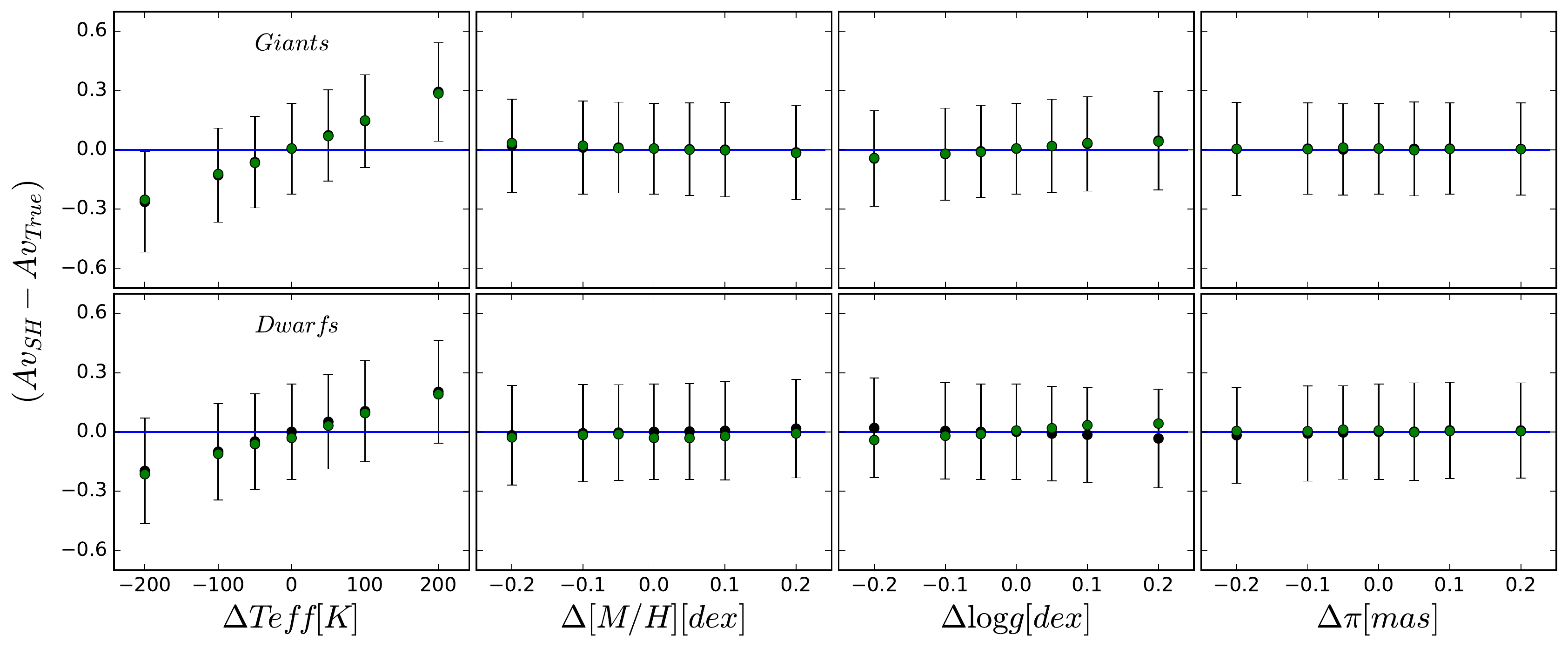}
  \caption{Relative errors per parameter (First two rows: $d$, third and fourth rows: $\tau$, fith and sixth rows: $m_{ast}$, seventh and eighth rows: $A_V$) obtained when different systematic offsets are introduced to the stellar parameter at a time. For this test we use the simulated PARSEC sample with high-resolution errors, but only 500 stars. The panels show the results for giant stars with $\log{g}<4$, and dwarf stars with $\log g>4$ separately. From the left, the first column shows the results for offsets to $T_{\mathrm{eff}}$, the second column shows the offsets to [M/H], the third column shows the offsets to $\log g$, and the last column shows the offsets to parallax $\pi$. In each panel, the black (green) dots represent the mean (median) of the distribution of relative errors in each parameter, whereas the standard deviation from the mean is shown as the error bars.} 
  \label{shiftparam}
\end{figure*}

As in \cite{Santiago2016}, we also tested the effect of systematic offsets in the observed quantities on our results, using the PARSEC samples. The results are shown in Figure \ref{shiftparam}. Again we split the simulated samples into giants and dwarfs. Each column corresponds to a given parameter for which shifts were applied, keeping the other parameters at their observed (but systematics-free) value. Each panel shows the mean, the median, and the dispersion around the mean values of the relative error (over all stars), as a function of the shift parameter.

In almost all panels, the difference between the mean and median relative errors is very small, attesting to the existence of relatively few outliers. The dispersion around the mean is rarely larger than 20\% (or 0.2 mag in the case of extinction) in most cases studied. The exception is the relative age error, for which the mean is much farther from zero than the median, and the dispersion is of order $\simeq 100\%$ in the case of giants, and even larger for dwarfs (see also discussion above). On average, however, the effect of systematic errors on our estimated parameters are typically less than $\pm 10\%$. In the following we discuss some more conspicuous effects.

Systematic errors in $T_{\mathrm{eff}}$ affect almost all inferred parameters for both dwarfs and giants. The effect of (under)overestimating temperatures on dwarfs is perhaps simpler to interpret, as it leads to best matching models of (lower)higher masses, therefore (less)more luminous. The apparent distance modulus is correspondingly biassed too (low)high, yielding either (smaller)larger inferred distances or $A_V$, or a mixture of both. For the giants, evolutionary timescales are shorter, making age a central parameter. In their case, an (under)overestimated $T_{\mathrm{eff}}$ requires a (older)younger (and therefore (less)more massive and luminous) progenitor, and leads to concordance models of higher apparent distance moduli.

The systematic effects of metallicity biases on the inferred parameters are of lower amplitude as compared to $T_{\mathrm{eff}}$. This is consistent with $T_{\mathrm{eff}}$ having a tighter correlation with the photometric parameters, and hence more strongly affecting the likelihood functions. Estimated masses are affected when the metallicity is biased. An (over)underestimated metallicity leads to a better match of a given star to models of (higher)lower metallicity, which will be of (higher)lower mass for a fixed luminosity (i.e., fixed apparent magnitudes and distance). For giants, the higher(lower) masses will again require younger(older) progenitors.

The case of $\log g$ is such that it affects distances more strongly. This has been investigated before by \cite{Santiago2016}, with similar results. An (under)overestimate in $log g$ leads to an (over)underestimate in the distances, since the data for a star become more consistent with models of stars (more)less luminous than it actually is. 

Parallaxes (last column of Figure \ref{shiftparam}) also predominantly affect distances, in the expected sense. A slight effect on masses and ages of giants can also be seen, since an (under)overestimated parallax will require (more)less luminous giants, therefore shifting the models towards (younger)older ages with (higher)lower mass progenitors.

\section{External Validation} \label{validation}

Up to this point, we have shown {\tt StarHorse} results for simulated stars, for which we previously know all their stellar parameters. Although TRILEGAL delivers realistic star-counts simulations for the Galaxy, real data can present different behaviour from the assumptions we made on the simulations. In this section we test {\tt StarHorse} results for observed data. We choose samples of stars from  eclipsing binaries, asteroseismology, and open clusters. Another important difference with respect to the validation carried out in the previous section is that here we will compare our estimated distances, ages, masses, and extinction values out to distances larger than 1 kpc.

\subsection{Eclipsing Binaries} \label{sec:binaries}
In this section we show {\tt StarHorse} validations with samples of Eclipsing Binaries (EBs). EBs can give precise stellar masses and radii, which in turn yield precise surface gravities. If temperature is also available, these stars can provide a good benchmark for estimating distances, ages, masses, and extinctions. Here we show tests on two samples: one is made up of detached binary systems with individual values of stellar parameters, the other has EBs that are photometricaly unresolved.   

\subsubsection{Detached Eclipsing Binaries}

We use the sample of EBs from \citet{Ghezzi2015} as a first comparison sample to our {\tt StarHorse} estimates. Those authors carried out a literature search for detached binary systems with at least one evolved star to be used as a benchmark for the determination of masses. This sample contains a total of 26 binaries with photometry, atmospheric parameters, parallaxes, ages, masses, and extinctions. The majority of the sample is composed of stars from the Large and Small Magellanic Clouds (LMC, SMC), but some stars are near the Solar vicinity ($d<1$ kpc).  As we see in Table \ref{sampletable}, this sample contains mainly giant and subgiant stars ($\log g <3.6$) with low metallicity. To run {\tt StarHorse} we used as input the photometry, $T_{\mathrm{eff}}$, [Fe/H], and $\pi$ given in Table 1 of \citet{Ghezzi2015}, and $\log g$ given in their Table 2. 

\begin{figure}
  \resizebox{\hsize}{!}{\includegraphics{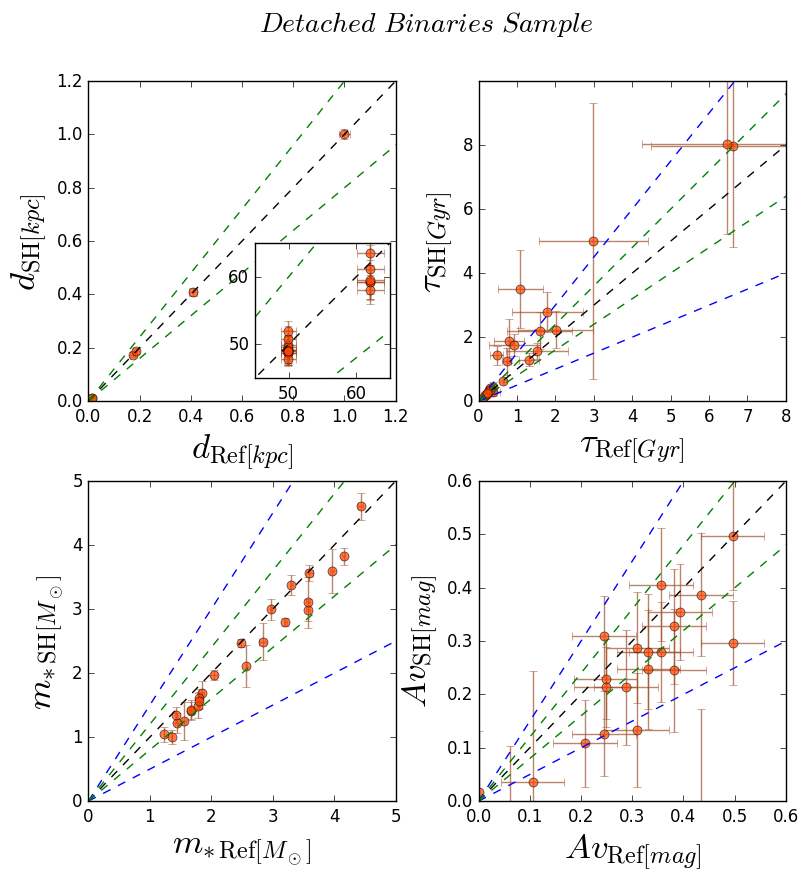}}
  \caption{Comparison between our distance (upper left panel), age (upper right panel), mass (lower left), and extinction (lower right) results with those from asteroseismology and the PARAM code for the sample of \citet{Ghezzi2015} binaries. The black dashed line is the identity line. The green dashed lines correspond to $\pm 20\%$ deviates, whereas the blue dashed lines (in the last three panels only) correspond to $\pm 50\%$. The insert on the bottom right shows the same comparison for the LMC/SMC stars.}
  \label{binaries}
\end{figure}

{\tt StarHorse} recovered distances, ages, masses and extinctions for 24 out of 26 stars. Fig. \ref{binaries} compares these results to the more fundamental determinations of \citet{Ghezzi2015}. For the distances, shown in the upper left panel of Fig. \ref{binaries}, the reference distance is the inverse of the {\it Hipparcos} input parallaxes used by those authors. Because the sample includes stars either in the Solar Neighbourhood, ($d < 1$ kpc), as well as stars at LMC/SMC distances, we show these latter as a separate inset in the figure. The agreement is excellent, with only a small degradation for the larger distances to LMC and SMC stars, whose parallax uncertainties are larger (see the panel inset). The mean uncertainty on the parallax of this sample is lower then the simulated samples; see Table \ref{sampletable}. This is probably the reason for such a good agreement in distances. In the same figure, we show the comparison between {\tt StarHorse} and PARAM \cite{daSilva2006} ages (which incorporate results from asteroseismology using a method similar to {\tt StarHorse}; upper right panel), and the comparison between {\tt StarHorse} masses and those taken from asteroseismology scaling relations (lower left panel; for the references on the masses, see Table 2 in \citealt{Ghezzi2015}). We also show the comparison between {\tt StarHorse} $A_V$ and $3.1E(B-V)$ (lower right panel), where the references for E(B-V) are in Table 1 of \cite{Ghezzi2015}.

From Figure \ref{binaries}, we see that {\tt StarHorse} yields ages that are systematically larger than those from PARAM. The median age offset is of $22\%$. Still, most of the {\tt StarHorse} ages appear to agree with PARAM ages to within 50\%. The two age determinations are also consistent with each other for most stars, considering both error bars. Table \ref{perctable} shows that 50\% (84\%) of the estimated ages have errors below 25\% (76\%). As for the masses, the agreement between our estimates and those from asteroseismology are somewhat better. {\tt StarHorse} masses tend to be smaller by $12\%$. The error bars and discrepancies relative to PARAM values are relatively smaller as well, with most of stars having errors of $17\%$ or smaller. The extinction estimates also agree well with the ones from the literature, despite the large error bars. Most of the stars have an $A_V$ error below $0.18$ mag, with a moderate systematic effect ($-0.07$ mag). We note that the systematics between StarHorse and the reference sample values cannot be due to systematics in the input atmospheric parameters, since both methods used the same data as input.

\subsubsection{Other Eclipsing Binaries}
As a second comparison with EBs, we use the sample from \citet{Stassun2016}, which contains 156 systems. Their sample is composed of stars with precise stellar radii and effective temperatures. Most of the stars have also available masses, Gaia parallaxes, metallicities, and magnitudes in at least one the following filter systems: Tycho \citep{Hog2000}, APASS \citep{Henden2014}, Strömgren \citep{Casagrande2014} and 2MASS \citep{Cutri2003}. Their distances and extinctions were estimated by performing fits to the broad-band photometric spectral energy distributions of the binary systems. The range of the parameters is described in Table \ref{sampletable}. Although most of the parameters are individual for each star, the magnitudes are systemic; these magnitudes describe the binary system, not each star individually. This can include a bias in our likelihood, and as most of the sample is made by systems with similar massas, this will affect both primary and secondary stars, though the effect would be grater on the secondary stars. It is very important to proceed with this test, since approximately 50\% of Solar-type field stars are binary or multiple systems \citep{Raghavan2010,Moe2017b}, and most of them are photometricaly unresolved.

\begin{figure}
  \resizebox{\hsize}{!}{\includegraphics{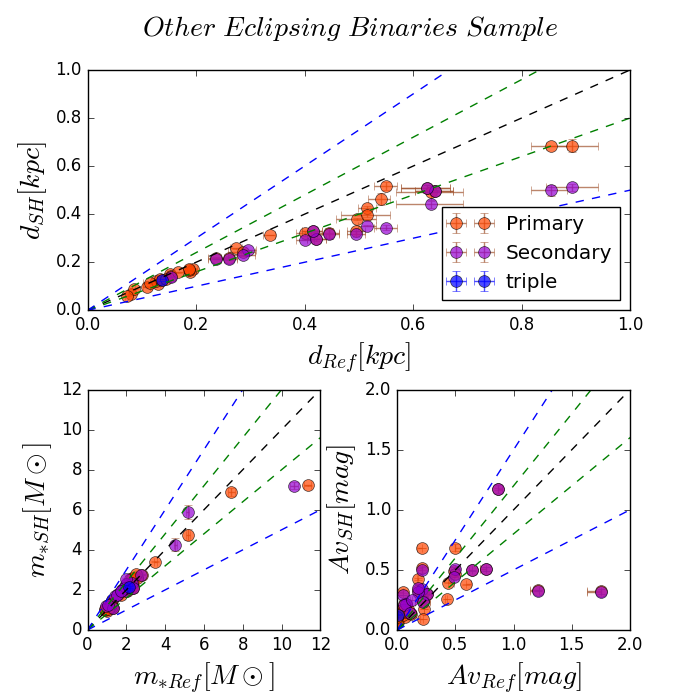}}
  \caption{Comparison between our distance (upper panel), mass (left lower panel) and extinction (right lower panel), results with those from the EB sample of \citet{Stassun2016}. The dashed lines correspond to the same deviates as in Figure \ref{binaries}. The legend shows the orange dots representing the primary star, the purple dots the secondary star, and the blue dots represents detected triple systems.}
  \label{StassunEB}
\end{figure}

To run {\tt StarHorse} we only select stars with available masses and limit the distance to 1 kpc. Figure \ref{StassunEB} compares the {\tt StarHorse} results for these stars with the ones of \citet{Stassun2016}. The {\tt StarHorse} distances are in general underestimated in relation to the ones estimated by \citet{Stassun2016}. Most of our distance estimates are smaller by 20\% (see Table \ref{perctable}), the effect being larger for the secondary members. This is probably a direct result of the systemic magnitudes we are using, since each star is assigned a brighter magnitude than it actually is, the amplitude of the effect being larger for secondaries. Therefore, our code will tend to match it with models at nearer distances. The masses estimated by {\tt StarHorse} are in reasonable agreement with the reference sample, as shown in Table \ref{perctable}: 50\% of the stars present relative errors smaller then 10\%. We should keep in mind that the surface gravities for this EBs sample are based on the quoted radii and masses. The estimated extinction values do not present systematic deviation with the extinction given by \citet{Stassun2016}, but as the extinction strongly depends on the distance models, they also should be affected by the systemic magnitudes. 

\subsection{Asteroseismology: CoRoGEE} \label{sec:corogee}

\begin{figure}
  \resizebox{\hsize}{!}{\includegraphics{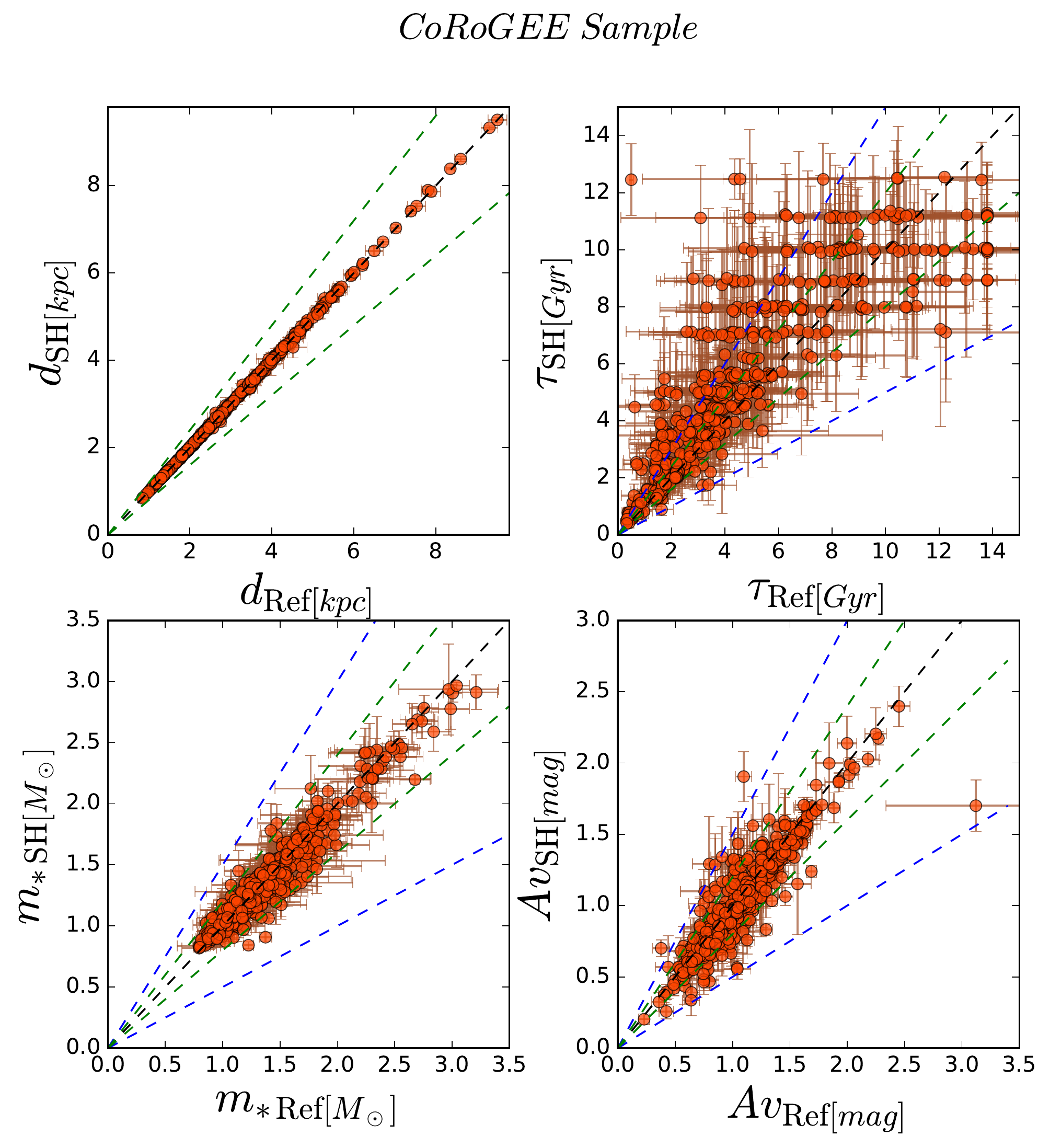}}
  \caption{Same panels as in Figure \ref{binaries}, but now showing the comparison to the CoRoGEE sample.}
  \label{corogee}
\end{figure}

We also use the CoRoT-APOGEE sample (CoRoGEE; \citealt{Anders2017}) to evaluate the accuracy of our {\tt StarHorse} results. CoRoGEE contains seismic measurements (from CoRoT) combined with high resolution spectra (from APOGEE) for more then 600 stars, this sample has also estimates of distance, age, mass, and $A_V$ from the PARAM code. They cover a wide range of Galactocentric distances, metallicities, and ages (see Table \ref{sampletable}), but are all red giants stars. To run {\tt StarHorse}, we use as input the atmospheric parameters and total metallicity given by APOGEE, and the parallaxes as the inverse of the distance given by PARAM. We then compare our estimates with the ones from PARAM. The comparison is shown in Figure \ref{corogee}. The estimated distances are in excellent agreement (upper left panel), since the well-constrained input parallaxes are used by {\tt StarHorse} when building the marginalized distance PDF for each star.

The upper right panel of Figure \ref{corogee} shows the comparison between {\tt StarHorse} and PARAM ages. The discreteness of our age grid is visible in the plot. The scatter is large, but most of the stars have ages that agree within $\pm 50\%$ of each other. The median and $84\%$-ile positions in the distribution of age discrepancies  are $12\%$ and $65\%$, respectively (Table \ref{perctable}). As in the case of the EBs, {\tt StarHorse} tends to yield larger ages than those based on asteroseimology, but this time with a smaller median systematic ($16\%$). In general, the results are quite similar to those from the previous section. This is also true for the extinction estimates, shown in the bottom right panel of Fig. \ref{corogee}. The mass estimates tend to show a better agreement in the case of CoRoGEE, with no systematic trend and smaller errors (median value of $4\%$) when compared to the EBs.

\subsection{The OCCASO Clusters} \label{sec:occaso}

As a third comparison sample, we use the data from the Open Clusters Chemical Abundances from Spanish Observatories (OCCASO) sample \citep{Casamiquela2016, Casamiquela2017}. This sample contains a total of 128 stars from 18 clusters, covering Galatocentric distances out to 6 kpc, and a small range in metallicity. OCCASO contains only red clump stars, for a better spectroscopic resolution in spectroscopy. The age and distance estimates for these clusters are based on isochrone fitting, with a typical uncertainty of 0.2 mag in distance modulus and a mean age uncertainty of 0.2 Gyr. The input parameters used for {\tt StarHorse} were parallaxes, converted from the isochrone distances, metallicities, and atmospherical parameters for each star from high-resolution spectroscopy from OCCASO survey; the mean uncertainty on these parameters can be seen in Table \ref{sampletable}.

\begin{figure}
  \resizebox{\hsize}{!}{\includegraphics{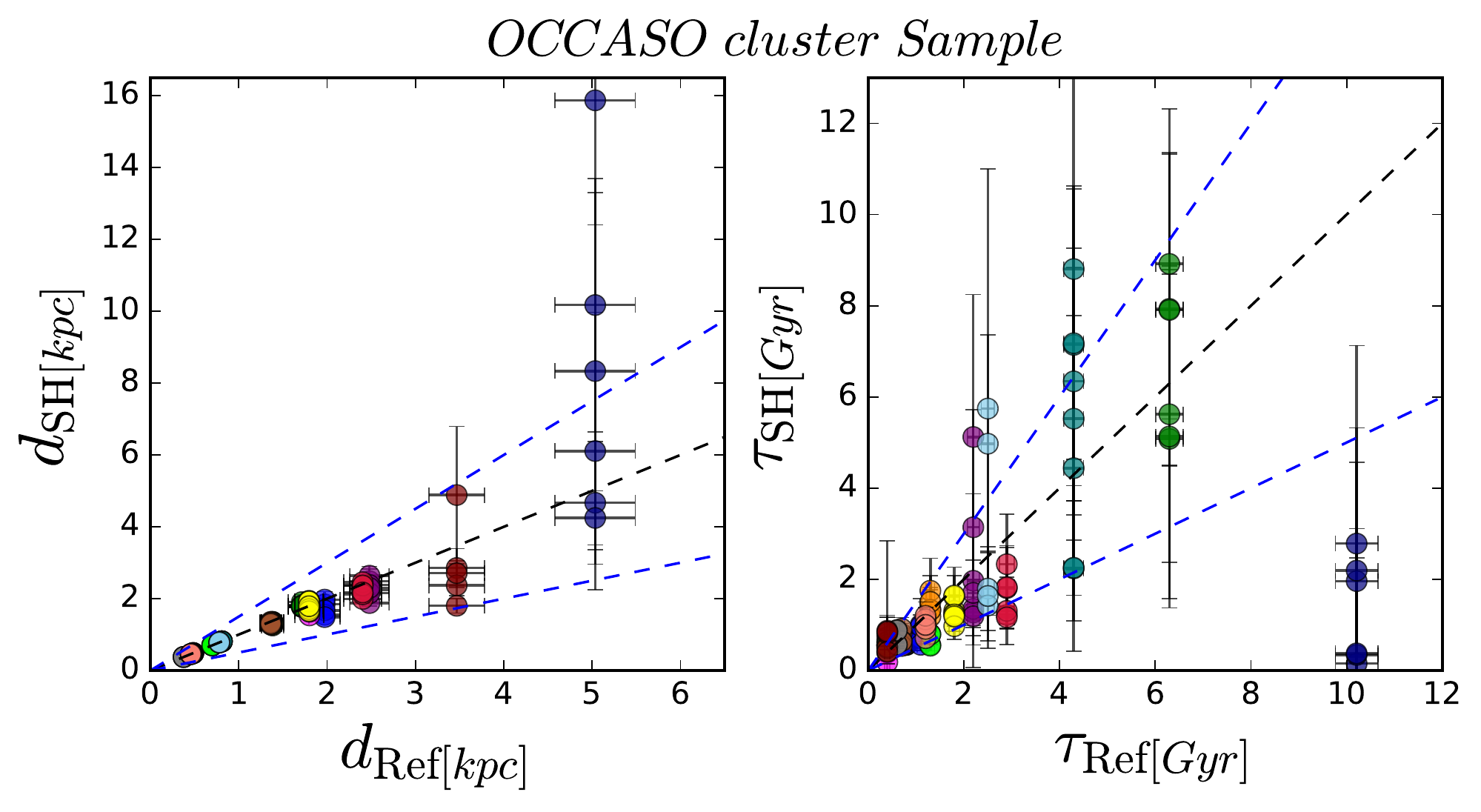}}
  \caption{Same as the upper panels in Figures \ref{binaries} and \ref{corogee}, but now showing the comparison of determined distances and ages to those for the star cluster sample from OCCASO, described in Sec. \ref{sec:occaso}. The stars in the figure are coloured as an identification of their cluster host. In this figure we show only the identity and $\pm 50\%$ deviating lines.}
  \label{occaso}
\end{figure}

Figure \ref{occaso} shows the comparison to {\tt StarHorse} distances and ages. Each point is a star, for which the cluster's distance and age are attributed. The distances agree with no systematics and median ($84\%$-ile) discrepancies of $2\%$ ($13\%$). Only a few of stars from the most distant cluster resulted with {\tt StarHorse} distances significantly above the one from isochrone fitting. The ages exhibit larger relative discrepancies ($31\%$ and $63\%$, respectively for the $50\%$-ile and $84\%$-ile positions). We notice from the figure that the stars  with a high discrepancy on distance also results in a high discrepancy in age (dark blue points), these points belong to the NGC 6791 cluster -- one of oldest open clusters in the Milky Way. Still, for most of the stars the age estimates are consistent within the relatively large error bars.

\begin{table*}
\centering
\caption{Results of the external validation tests. We report the percentiles of the modulus of the relative deviation distribution and the median of the relative deviation distribution for each parameter estimated by {\tt StarHorse}. Data are shown for all reference samples used.}
\setlength{\tabcolsep}{2pt}
\begin{tabular}{cccccccccc}
\hline
\hline
Sample & Parameter  & $P_5$ & $P_{16}$  & $P_{50}$ & $P_{84}$ & $P_{95}$ & median  \\
\hline 
PARSEC          & $d$   & 0.005 & 0.015 & 0.063 & 0.163 & 0.258 & 0.0 \\
high-res        & $\tau$        & 0.002 & 0.109 & 0.350 & 0.972 & 4.203 & 0.0 \\
                & $m_{\ast}$      & 0.001 & 0.012 & 0.074 & 0.222 & 0.391 & 0.0 \\
                & $Av$        & 0.005 & 0.014 & 0.048 & 0.107 & 0.162 & 0.0 \\
\hline
PARSEC          & $d$  & 0.005 & 0.016  & 0.060 & 0.172 & 0.289 & -0.001\\
low-res         & $\tau$       & 0.014 & 0.110  & 0.398 & 1.120 & 4.550 & 0.002\\
                & $m_{\ast}$      & 0.003 & 0.017  & 0.083 & 0.247 & 0.429 & -0.001\\
                & $Av$        & 0.016 & 0.049  & 0.169 & 0.354 & 0.499 & 0.004\\
\hline
TRILEGAL             & $d$  & 0.007 & 0.023 & 0.079 & 0.171 & 0.240 & -0.04 \\
 all priors          & $\tau$       & 0.017 & 0.054 & 0.188 & 0.516 & 0.937 & 0.11 \\
                     & $m_{\ast}$      & 0.006 & 0.019 & 0.063 & 0.132 & 0.206 & -0.04 \\
                     & $Av$        &  0.003 & 0.012 & 0.043 & 0.099 &  0.162 & 0.007 \\
\hline
Detached            & $d$  & 0.001 & 0.002 & 0.017 & 0.041 & 0.047 & -0.015 \\
Eclipsing Binaries  & $\tau$       & 0.009 & 0.022 & 0.246 & 0.764 & 1.94 & 0.22\\
                   & $m_{\ast}$      & 0.008 & 0.033 & 0.124 & 0.165 & 0.194 & -0.12 \\
                   & $Av$        & 0.017 & 0.030 & 0.072 & 0.186 & 0.354 & -0.07\\
\hline
Other                    & $d$  & 0.032 & 0.071 & 0.178 & 0.298 & 0.419 & -0.178 \\
Eclipsing Binaries        & $m_{\ast}$  & 0.005 & 0.021 & 0.099 & 0.168 & 0.23 & 0.048 \\
                        & $Av$        & 0.021& 0.044 & 0.130 & 0.269 & 0.338 & 0.075\\
\hline
CoRoGEE    & $d$  & 0.000 & 0.001 & 0.003 & 0.008  & 0.014 & 0.0  \\
                & $\tau$       & 0.013 & 0.047  & 0.120 & 0.650 & 1.253 & 0.16 \\
                & $m_{\ast}$      & 0.003 & 0.011 & 0.043 & 0.097 & 0.164 &0.0  \\
                & $Av$        & 0.009 & 0.024 & 0.070 & 0.158 & 0.270 & -0.04\\
\hline
OCCASO                              & $d$  & 0.000 & 0.001 & 0.0244 & 0.130 & 0.310 & 0.0 \\
cluster members                     & $\tau$       & 0.028 & 0.090 & 0.310  & 0.626 & 0.968 & -0.07 \\
\hline
\end{tabular}
\label{perctable}
\end{table*}

\section{Released data products}\label{dataproducts}

We applied {\tt StarHorse} to a few spectroscopic surveys together with astrometric and photometric measurements to deliver public distances and extinction catalogues. For the moment we are not releasing ages and masses because their estimates are still subject to considerable improvement. Among the several spectroscopic surveys available to the community, we chose among those that have been more widely used and which have high to medium resolution. 
\noindent All catalogues are available to the community via the LineA web page\footnote{\url{http://www.linea.gov.br/020-data-center/acesso-a-dados-3/spectrophotometric-distances-starhorse-code/}\label{fot:url}}. As our method is subject to further improvements (e.g. incorporating new Gaia information, new stellar tracks with denser model grids, future releases of the adopted databases, and new extinction priors), this will lead to new versions of the released data, therefore, we suggest the reader to check for next releases at the LineA web page.
 
The following subsections explain the procedure used to determine distances and extinctions for each spectroscopic survey. We applied {\tt StarHorse} in these samples with all priors described in \S \ref{sec:updates}, and in all cases the following corrections were made to each survey catalogue:

\begin{enumerate}
\item {\it Polish photometry} \\
  In some few cases in which a 2MASS \citep{Cutri2003} magnitude of a star exists, but the associated uncertainty is -9999.99, we substitute this value by 0.2 mag. Similarly, if the quoted APASS magnitude uncertainty is 0, we set it to be 0.15 mag, and introduce an error floor of 0.02 mag. \\
\item {\it Correct metallicities for [$\alpha$/Fe]-enhancement} \\
  Since the PARSEC 1.2S stellar models do not yet include models with non-Solar [$\alpha$/Fe] ratio, we correct for this effect in the data, following \citet{Salaris1993}, by defining a total metallicity [M/H] as:
  $${\rm [M/H] = [Fe/H]} + \log(0.638\cdot[\alpha/{\rm Fe}]^{10} + 0.362)$$
  $${\rm \sigma_{\rm [M/H]} = \sqrt{\sigma_{\rm [Fe/H]}^2+\sigma_{\rm [\alpha/Fe]}^2}}$$
\end{enumerate}

\begin{table*}
\centering
\caption{General description of the release distances and extinctions}
\setlength{\tabcolsep}{2pt}
\begin{tabular}{ccc}
\hline
\hline
Column & Description & units \\
OBJECT ID & Survey's object ID name \\
glon & Galactic longitude & degrees \\ 
glat & Glactic latitude & degrees \\
dist05 & 5th percentile of the stars's distance PDF & kpc\\
dist16 & 16th percentile of the stars's distance PDF & kpc\\
dist50 & 50th percentile of the stars's distance PDF & kpc\\
dist84 & 84th percentile of the stars's distance PDF & kpc\\
dist95 & 95th percentile of the stars's distance PDF & kpc\\
meandist & Mean of the stars's distance PDF & kpc \\
diststd & standard deviation of the stars's distance PDF & kpc \\ 
AV05 & 5th percentile of the stars's extinction PDF & mag\\
AV16 & 16th percentile of the stars's extinction PDF & mag\\
AV50 & 50th percentile of the stars's extinction PDF & mag\\
AV84 & 84th percentile of the stars's extinction PDF & mag\\
AV95 & 95th percentile of the stars's extinction PDF & mag\\
meanAV & Mean of the star's extinction PDF & mag\\
stdAV & Standard deviation of the stars's extinction PDF & mag\\
SH\_INPUTFLAGS & {\tt StarHorse} flags regarding the input data & \\
SH\_OUTPUTFLAGS & {\tt StarHorse} flags regarding the output data & \\
\hline
\end{tabular}
\label{pubrelease}
\end{table*}

\subsection{APOGEE Catalogues}\label{apoapp}

The APOGEE-2 survey \citep{Majewski2017} is a program from the Sloan Digital Sky Survey (SDSS-IV, \citealt{Blanton2017}).  It is a spectroscopic survey conducted in the near infrared, with high resolution ($R\sim 20,500$), and high signal-to-noise ($S/N > 100$). It maps the Galaxy through all populations and it has targeted, as of its latest release (DR14; \citealt{Abolfathi2017}), about 270,000 stellar spectra. As a near infrared survey of high resolution, APOGEE has the advantage to study with more detail the stars located in the dusty regions of our Galaxy, such as the Bulge and Disk. We applied StarHorse to the latest APOGEE release, DR14.

\subsubsection{APOGEE DR14 ASPCAP}

We used the APOGEE DR14 allStar summary catalogue ({\tt allStar-l31c.2.fits}\footnote{available at \url{https://data.sdss.org/sas/dr14/apogee/spectro/redux/r8/allStar-l31c.2.fits}}). The file contains photometry in the 2MASS $JHK_s$ passbands \citep{Cutri2003}, as well as chemical abundances and atmospheric parameters derived by the APOGEE Stellar Parameters and Chemical Abundances (ASPCAP; \citealt{GarciaPerez2016}). Starting from this file we took the following pre-processing steps before running {\tt StarHorse}:
\begin{enumerate}
\item {\it Cross-match with photometry and astrometry}\\
  Using TOPCAT \citep{Taylor2005}, we did a positional crossmatch so as to add information from APASS DR9 \citep{Henden2014} (207,604 matches) and {\it Gaia} DR1/TGAS \citep{GaiaCollaboration2016} (46,033 matches). Our crossmatch used a maximum separation of 5 arcsec. 
\item {\it Selecting reliable results}\\
  We discarded sources with ASPCAP flags containing the words ''BAD'' or ''NO$\_$ASPCAP''. In addition, we selected only sources with valid 2MASS $H$ magnitudes. This resulted in a total of 226,323 stars.
\item {\it Using uncalibrated ASPCAP results}\\
  When no calibrated ASPCAP results for one or more spectroscopic parameters are available (stars outside the calibration ranges; e.g., $\log g$ values for dwarfs), we use uncalibrated ASPCAP results. For those objects we inflated the uncertainties. Our ad-hoc conservative uncertainty estimates for these cases amount to 150 K in $T_{\mathrm{eff}}$, 0.3 dex in $\log g$, 0.15 dex in [M/H], and 0.1 dex in [$\alpha$/M].
\item {\it $A_V$ prior}\\
  As priors for the $V$-band extinction we used the APOGEE targeting extinction values \citep{Zasowski2013}, derived by the RJCE method \citep{Majewski2011}, by setting $A_{V, {\rm prior}}=A(K_s)_{\rm Targ} / 0.12$. When RJCE estimates were not available, we used $E(B-V)$ estimates from \citep{Schlegel1998} to estimate the prior $A_V$. As explained in Sec. \ref{ext}, the posterior $A_V$ values are allowed to lie in a very broad interval around the prior values. 
\end{enumerate}

\begin{figure*}
\centering
  \includegraphics[width=.8\textwidth]{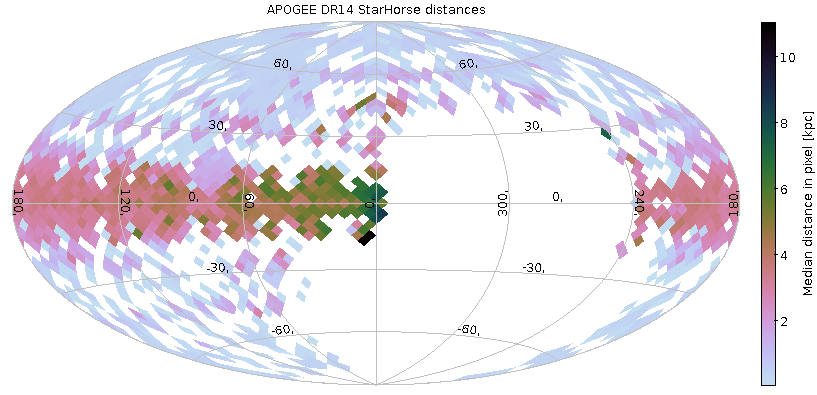} \\
  \includegraphics[width=.8\textwidth]{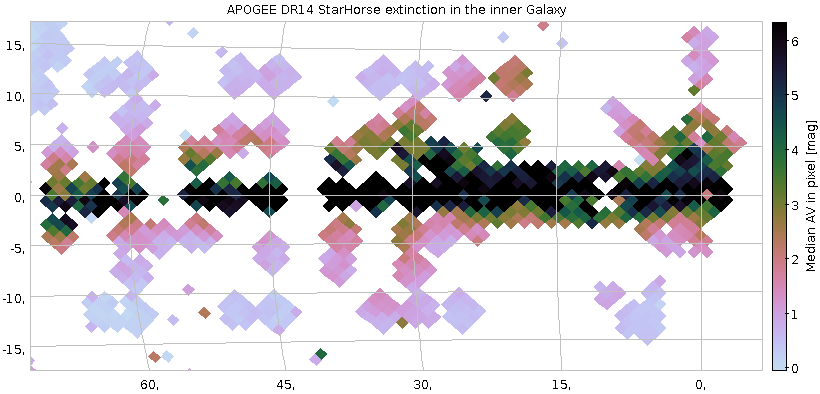} \\
  \includegraphics[width=.45\textwidth]{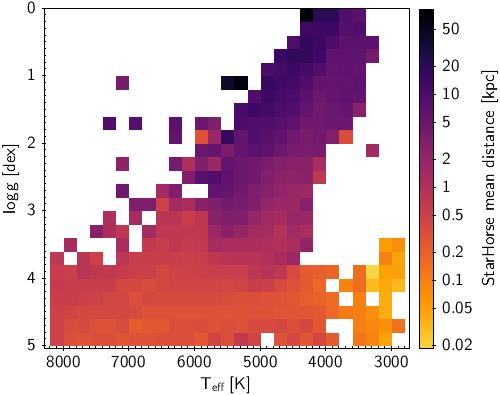} 
  \includegraphics[width=.45\textwidth]{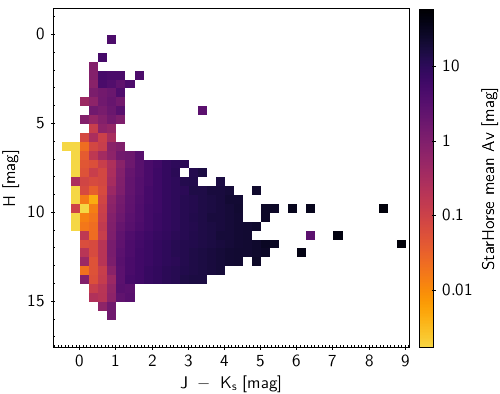} \\
  \caption{Illustration of the APOGEE DR14 distance and extinction results from {\tt StarHorse}. Top panel: Aitoff projection of median APOGEE distances per HealPix cell in Galactic coordinates. Middle panel: Resulting median $A_V$ per HealPix cell in the inner Galaxy. Bottom left panel: Spectroscopic Hertzsprung-Russell diagram, colour-coded by median distance in each pixel. Bottom right panel: 2MASS colour-magnitude diagram, colour-coded by median extinction in each pixel.}
  \label{apogeesum}
\end{figure*}

\begin{figure}
  \includegraphics[width=8.4cm, trim=0cm 0cm 0cm 0cm, clip=true]{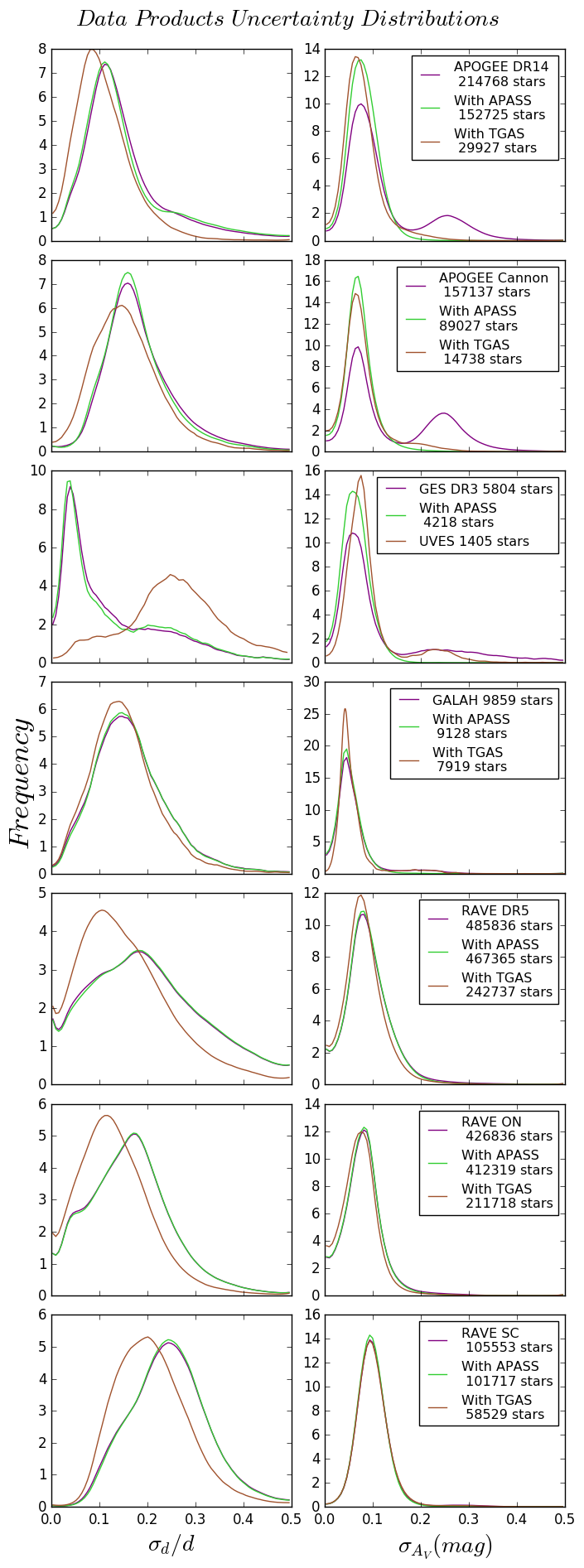}
    
  \caption{{\tt StarHorse} relative distance (left panels) and absolute extinction (right panels) uncertainty distributions for all release data products. The legend specifies the stars with available APASS photometry, parallaxes from TGAS or UVES spectroscopy (GES sample).}
  \label{vacsuncert}
\end{figure}

We then applied {\tt StarHorse} to the resulting catalogue. Our results are available through an SDSS-IV value-added catalogue (VAC) of APOGEE stellar distances, as part of SDSS DR14\footnote{\url{http://www.sdss.org/dr14/irspec/spectro_data/}} and on the Line web-page. The output format of this catalogue is described in Table \ref{pubrelease}. Figure \ref{apogeesum} summarises the results for APOGEE DR14. Figure \ref{vacsuncert} shows the associated uncertainty distributions.

\subsubsection{APOGEE DR14 Cannon}
As the APOGEE DR14 also contains stellar parameters and chemical abundances derived by the data driven method called Cannon \citep{Ness2015,Casey2016a}, we also applied StarHorse with this input. We used the APOGEE DR14\footnote{\url{http://www.sdss.org/dr14/irspec/spectro_data/}} allStar summary catalogue ({\tt allStarCannon-l31c.2.fits}). Which contains the atmospheric stellar parameters, chemical abundances and its uncertainties. Starting from this file we did the following steps to prepare a input to {\tt StarHorse}:\\
\begin{enumerate}
\item {\it Cross-match with photometry and astrometry}\\
  We carried out a positional cross match, using TOPCAT \citep{Taylor2005}, with photometry from APASS \citep{Henden2014} and from 2MASS \citep{Cutri2003}.
\item {\it Av Prior}\\
   As priors for the $V$-band extinction we used the APOGEE targeting extinction values \citep{Zasowski2013}, derived by the RJCE method \citep{Majewski2011}, by setting $A_{V, {\rm prior}}=A(K_s)_{\rm Targ} / 0.12$.
\end{enumerate}

\subsection{The Gaia-ESO sample} \label{GES}
The {\it Gaia}-ESO survey \citep[GES][]{Gilmore2012a} is a large public spectroscopic survey with high resolution that covers all Milky Way components and open star clusters of all ages and masses. The final GES release is expected to include about $10^5$ stars.

We downloaded the {\it Gaia}-ESO data release 3 (DR3) from the ESO catalogue facility\footnote{\url{http://www.eso.org/rm/api/v1/public/releaseDescriptions/91}}. This catalogue contains a total of 25533 stars, including the Milky-Way field, open clusters, and calibration stars. The catalogue contains 100 columns, from which we selected the necessary information to run {\tt StarHorse}: atmospheric parameters, elemental abundances, and 2MASS $JHK_s$ pass-bands \citep{Cutri2003}. From this catalogue we proceeded with the following steps, before applying {\tt StarHorse}. \\

\begin{enumerate}
\item{\it{Select only field stars and reliable sources:}} \\
From this catalogue we selected only Milky Way field stars, which are the main targets intended for {\tt StarHorse}. This leaves us with 7870 stars. We then adopted the following quality criteria: relative errors in Teff less than 5\%; errors in logg lower than 0.4 dex; and  errors in metallicity lower than 0.2 dex, with no cuts in the abundances. From the 7870 field stars, 6316 of them meet these criteria. \\
\item{\it{Calculate an overall $[\alpha/Fe]$ abundance}:}\\
The PARSEC models do not list individual elemental abundances, only the total metallicity value, [M/H]. The models do not include non-Solar abundances in $\alpha$ elements either. Therefore, we converted the GES abundances in the total metallicity [M/H] before running {\tt StarHorse}. For that purpose, we calculated the overall [$\alpha$/Fe] abundance as follows:
  $$
  {\rm [\alpha/Fe]} = \frac{1}{n} \sum\limits_{i=1}^n {\rm [X_i/H] - [Fe/H]},
  $$  
where $X_i$ refers to the elements O, S, Ti, Ca and Mg, and the [X$_i$/H] abundances were calculated using the solar values from \citet{Asplund2009}. The error in [$\alpha$/Fe] was propagated in quadrature from the error on each elemental abundance and the error in [Fe/H]. \\

\item {\it{Cross-match with photometry}}: \\
To obtain more precise and reliable extinction estimates we decided to include APASS \citep{Henden2014} magnitudes for this sample. For this we carried out a positional crossmatch with APASS DR9 \citep{Henden2014} using TOPCAT \citep{Taylor2005}, with a maximum sparation of 1 arcsec. Of the 6316 stars, 5719 stars have APASS magnitudes.\\

\item{$A_V$ Prior:}\\
  No $A_V$prior was applied to this sample, therefore the posterior probability function for $A_V$ always ranges from 0 to 3 (mag).\\ 

 \end{enumerate}
  After carrying out these steps, we used this final catalog with 6316 stars as input to {\tt StarHorse}. The code delivered an output catalog with 6011 stars with available distances and extinctions. The columns of the released distances and extinction catalogue are shown in Table \ref{pubrelease}, and the uncertainties distribution are shown in Figure \ref{vacsuncert}.
  
\subsection{GALAH sample} \label{galah}

The Galactic Archaeology with HERMES (GALAH; \citealt{Martell2017}) is a spectroscopic survey that will target about 1 million stars with the high-resolution ($R\sim 28,000$) instrument High Efficiency and Resolution Multi-Element Spectrograph (HERMES; \citealt{deSilva2012}), at the Anglo-Australian Telescope (AAT). The main goal of the project is to provide a detailed star-formation history for the thick and thin disks. Therefore, the survey covers mainly the disk, but it also has some fields towards the bulge and halo. A first public data release of GALAH is already available. We then applied {\tt StarHorse} to estimate distances and extinction for GALAH DR1 stars. The catalogue\footnote{ \url{https://cloudstor.aarnet.edu.au/plus/index.php/s/OMc9QWGG1koAK2D}} contains 10680 stars with [Fe/H], [$\alpha$/Fe], $\log$ g, and $T_{\rm eff}$ measurements derived by the Cannon method \citep{Ness2015}. We proceeded with the following steps to have an input catalogue ready for {\tt StarHorse}:
\begin{enumerate}
\item{\it{Stellar parameter uncertainties}} \\
  The GALAH DR1 catalogue does not provide individual uncertainties for the stars. We therefore used the values recommended by \cite{Martell2017}: $\sigma$([Fe/H]) = 0.056 dex, $\sigma(\log g) = 0.17$ dex and $\sigma(T_{\rm eff}) = 51$ K. As there is no mention of the uncertainty in [$\alpha$/Fe], so we assumed the error to be same as for $\sigma$[Fe/H].\\

\item {\it{Cross-match with photometry and astrometry}}: \\
  Since the GALAH DR1 catalog has stars in commom with the {\it Gaia} DR1, we carried out a positional crossmatch with {\it Gaia} using TOPCAT \citep{Taylor2005}, with a maximum sparation of 1 arcsec. From the 10680 DR1 stars, 7919 stars have parallax available. We obtained photometry for the sample by crossmatching with the 2MASS \citep{Cutri2003} (10680 matches) and APASS \citep{Henden2014} (9263 matches) catalogues.\\
\item {\it{Av Prior}} \\
  We use the reddening given by the GALAH DR1 catalogue, which is derived by comparison of absolute magnitudes with the apparent magnitude $V$ from APASS \citep{Henden2014} and $J$ magnitude from 2MASS\citep{cutri2013}. We assumed then that $A_V = 3.1E(B-V)$.
  
\end{enumerate}
After these steps were completed, we used this final file as input to {\tt StarHorse} and derived distances and extinctions. The code returned 10,623 distances and extinctions. The columns for the released catalogue are described in Table \ref{pubrelease} and the uncertainty distribution is shown in Figure \ref{vacsuncert}.  

\subsection{The RAVE catalogues}\label{raveapp}

The RAdial Velocity Experiment (RAVE, \citealt{Steinmetz2006}) is one of the largest spectroscopic surveys of the Milky Way. RAVE has already delivered spectra for almost 500K stars, that were randonly targeted in an area of 20K square degrees of the Galactic Southern Hemisphere. In addition, RAVE is currently the survey that contains the largest number of stars in common with TGAS ($\sim$ 200K stars). The survey works with a multi-object spectrograph deployed on 1.2-m UK Schmidt Telescope of the Australian Astronomical Observatory (AAO). The spectra have a medium resolution of ($R\sim 7,500$) and cover the Ca-triplet region (8410-8795 \textup{\AA}). Given that RAVE has a medium resolution, is pioneering among the large spectroscopic surveys, covers a large area, and has the largest overlap with the Gaia sample, we decided to apply {\tt StarHorse} to the entire survey, and to make the estimated distances and extinctions available to the community. The following subsections explain how we proceed with the RAVE input catalogues to execute {\tt StarHorse} and the description of the released distances-extinction catalogues. All RAVE catalogues were downloaded from the RAVE website\footnote{\label{note1}\url{https://www.rave-survey.org/project/}}.

\subsubsection{The RAVE DR5 catalogue}\label{ravedr5}
The data release 5 \citep[DR5][]{Kunder2017} is the lastest RAVE data release. It contains spectra for 483,330 stars. We downloaded the publicly released catalogue called RAVE\_DR5, which contains spectral parameters and radial velocities derived by the SPARV pipeline \citep{Zwitter2008,Siebert2011}. The catalogue also contains astrometry from {\it Gaia}-DR1 (215,590), and photometry from 2MASS and APASS. We note that very recently, \cite{McMillan2017} updated the DR5 catalogue parameters and derived distances using feedback from the Gaia DR1 parallaxes, but our VAC presented here is based on the public DR5 data. From this catalogue, we proceeded with the following steps before running {\tt StarHorse}:
\begin{enumerate}
\item{\it{Spectral parameters}}\\
 We use the calibrated atmospheric parameters, which are named in the catalogue as Teff$_{NK}$, logg$_{NK}$, Met$_{NK}$. For the uncertainties, if the error spectral analysis is available, we use the maximum between the two values: $\sigma$Teff$_{K}$ and StdDevTeff$_K$, otherwise we use the maximum between: 70K and $\sigma$ Teff$_K$. We worked analogously with the other parameters.
\item{\it{ $A_V$ prior}}\\
  As explained in section \ref{ext}, we can use a prior value of extinction to build the $A_V$ posterior probability function. We use as extinction prior in $V$-band the maps of $E(B-v)$ from \cite{Schlegel1998} for this catalogue. 
  
\end{enumerate}

\subsubsection{RAVE-SC catalogue}\label{ravesc}
The RAVE-SC catalogue has stars from DR5 with gravity from seismic calibrations \citep{Valentini2017}, therefore this sample is only compose by giants. We downloaded the catalogue named as RAVE\_Gravity\_SC. Step 1 from the previous subsection \ref{ravedr5} was also applied to this sample. We use the overall [$\alpha$/Fe] abundance and the [Fe/H], given by the catalogue to calculate a total metallicity as defined by \citep{Salaris1993}, with a fixed uncertainty of 0.2 dex. The atmospheric parameters were used as they were given by the catalogue; we use the following temperature and surface gravity columns: Teff\_IR and logg\_SC. For the $A_V$ prior we use the \cite{Schlegel1998} $E(B-V)$ maps.

\subsubsection{RAVE-on catalogue}\label{raveon}
The RAVE-on catalogue \citep{Casey2017} has stars from DR5 with parameters derived by the Cannon method \citep{Casey2016a}. We downloaded the catalogue named as RAVE-ON. The atmospheric parameters and [Fe/H] were used directly from this catalogue. The following steps were applied before applying {\tt StarHorse}:
\begin{enumerate}
\item{\it{Calculate an overall $[\alpha/Fe]$ abundance}}\\
  The Cannon provides the individual abundances for the stars. We then calculated [$\alpha$/Fe] as the simple average between the individual abundances when they are available, exactly as described in section \ref{GES}, with $X_i$ as O, Mg, Ca, and Si. 
\item{$A_V$ Prior:}\\
  No $A_V$prior was applied to this sample, therefore the posterior probability function for $A_V$ always ranges from 0 to 3 (mag).\\   
\end{enumerate}

\subsection{{\tt StarHorse} FLAGS}
All released data products catalogues have two columns that describe the {\tt StarHorse} input data, SH\_INPUTFLAGS, and the {\tt StarHorse} output data, SH\_OUTPUTFLAGS, as shown in Table \ref{pubrelease}. The input flags specify which parameters were used in the likelihood calculation to estimate the distances and extinctions given. For example, if the temperature was available for that star a TEFF flag will appear, and if TEFF was not available in the calculation a uncalTEFF flag will appear. The other parameters are specified as follows in the input flag: LOGG (surface gravity), PARALLAX (parallax), MH (metallicity), JHKs (2Mass filters) and BVgri (APASS filters). If the input flag contains ALPHAM it means that the alpha elements were available in the calculation of the total metallicity of the star. The input flags also indicate you if we use a $A_V$ prior as the AVprior flag. The output flags tell us if the number of models which have converged in the likelihood calculation is too small. If less then 10 models are consistent with the star a NUMMODELS\_BAD flag will appear, while if the number of models is between 10 and 30 a NUMMODELS\_WARM flag will appear. The output flags also indicate if the estimated extinction is negative (NEGATIVE\_EXTINCTION\_WARN), if it is too high (HIGH\_EXTINCTION\_WARN), or if the estimated extinction has a bright 2Mass source (EXTINCTION\_BAD\_BRIGHT2MASS). \\\\

\section {Summary and Future Perspectives}\label{conclusion}

We have presented a code that computes distances, ages, masses, and extinctions for field stars with photometric, spectroscopic, and astrometric data. It is based on Bayesian inference, computing the marginal posterior distributions for the data given a set of stellar models. The code represents a significant improvement over the one presented by \cite{Santiago2016} in several aspects. The most important one is the ability to estimate ages, masses, and extinction, in addition to the spectrophotometric distances presented by those authors. The updated code, which we call {\tt StarHorse}, is also capable of incorporating the parallax as an additional observational quantity in the statistical analysis (Figure \ref{SHflux}). Updated spatial, metallicity, and age priors for the Galactic components, now including the bulge as well, are presented (Table \ref{priorgauss}). In addition, {\tt StarHorse} is now more flexible in terms of the input data and the choice of observational quantities to be used within them.

The new code was validated using simulated and real stars. These latter are samples with reliable parallax (or distance) data, including field giants with asteroseismic data, EBs used as benchmarks for stellar evolutionary codes, or cluster stars with well-known distances and ages, usually from isochrone fitting, often in combination with spectroscopic data. For EBs that are not detached, the distances present an offset in relation to the reference ones, our distances being usually smaller by 20\% in this specific case. The discrepancy is larger for secondary stars than for the primary ones, which is what one expects from using systemic photometric measurements. In all cases, age is the single most difficult parameter to infer, yielding median errors that range from $12\%$ to $35\%$ for quality spectrophotometric data, depending on the sample (Table \ref{perctable}). Errors larger than 100\% in age may result for $\simeq 15\%$ of the stellar models, most of them younger than $\tau \simeq 1$ Gyr. In a realistic flux-limited sample, as simulated by the TRILEGAL code or for real stars, the fraction of such catastrophic age errors is reduced to $\simeq 5\%$ of the stars. Our results for stellar ages, either based on simulated or real stars, also indicate a systematic trend of {\tt StarHorse} overestimating ages by $10-20\%$.

As for spectrophotometric masses, we obtain consistent results over all validation samples used, in the sense that errors $< 20\%$ are observed for most ($84\%$) of the stars in any sample. The median error varies depending on the quality of the parallax used as a constraint. For typical {\it Gaia}-TGAS errors of $0.3$ mas, the typical distance errors are around $15\%$. For real stars used as reference, median {\tt StarHorse} $A_V$ errors are of $0.07$ mag, with the 84$\%-ile$ error at $0.15-0.20$ mag. For TRILEGAL and PARSEC synthetic stars, the relatice errors are a bit smaller, around $0.04$ mag.  

We note that the error estimates based on comparison to real samples may be overestimated, considering that the some of the discrepancy may be attributed to the methods used to obtain the reference quantities for comparison. In fact, \cite{Rodrigues2017} report that ages and masses from asteroseismology are typically obtained with a precision of $19\%$ and $5\%$, respectively, which are comparable to the errors we quote in this analysis.

{\tt StarHorse} has already been used to infer distances and extinction values for stars from APOGEE DR14. These parameters, in turn, may be used in connection to APOGEE abundances and radial velocities, to study the properties of the main Galactic populations, and their spatial variations, as was previously done by \cite{Anders2014}, and \cite{Fernandez-Alvar2016} using distances from \cite{Santiago2016}. For more local samples, such as {\it Gaia}-TGAS and RAVE, reliable parallax information can be included in the Bayesian method to yield masses and ages, as validated in this paper, allowing for a more detailed modelling of the chemo-dynamical history of our Galaxy \citep{Anders2017}.

{ Finally, we have run {\tt StarHorse} on different public catalogs from the RAVE collaboration, as well as on GES, and GALAH public data releases. These are available for download at the LIneA web site.}\footnote{\url{http://www.linea.gov.br/020-data-center/acesso-a-dados-3/spectrophotometric-distances-starhorse-code/}}

\section*{Acknowledgements}
\addcontentsline{toc}{section}{Acknowledgements}

The {\tt StarHorse} code is written in python 2.7 and makes use of several community-developed python packages, among them {\tt astropy} \citep{AstropyCollaboration2013}, {\tt ezpadova}\footnote{\url{https://github.com/mfouesneau/ezpadova}}, {\tt numpy} and {\tt scipy} \citep{Oliphant2007}, and {\tt matplotlib} \citep{Hunter2007}. The code also makes use of the photometric filter database of VOSA \citep{Bayo2008}, developed under the Spanish Virtual Observatory project supported from the Spanish MICINN through grant AyA2011-24052.

We thank Eddie Schlafly (LBL) as well as the anonymous referee for various useful comments that helped improving the manuscript.

We thank Laia Casamiquela for providing data from the OCCASO survey. 

Funding for the SDSS Brazilian Participation Group has been provided by the 
Minist\'erio de Ci\^encia e Tecnologia (MCT), Funda\c{c}\~ao Carlos Chagas 
Filho de Amparo \`a Pesquisa do Estado do Rio de Janeiro (FAPERJ), Conselho 
Nacional de Desenvolvimento Cient\'{\i}fico e Tecnol\'ogico (CNPq), and 
Financiadora de Estudos e Projetos (FINEP).

Funding for the Sloan Digital Sky Survey IV has been provided by
the Alfred P. Sloan Foundation, the U.S. Department of Energy Office of
Science, and the Participating Institutions. SDSS-IV acknowledges
support and resources from the Center for High-Performance Computing at
the University of Utah. The SDSS web site is \url{www.sdss.org}.\\

SDSS-IV is managed by the Astrophysical Research Consortium for the 
Participating Institutions of the SDSS Collaboration including the 
Brazilian Participation Group, the Carnegie Institution for Science, 
Carnegie Mellon University, the Chilean Participation Group, the French Participation Group, Harvard-Smithsonian Center for Astrophysics, 
Instituto de Astrof\'isica de Canarias, The Johns Hopkins University, 
Kavli Institute for the Physics and Mathematics of the Universe (IPMU) / 
University of Tokyo, Lawrence Berkeley National Laboratory, 
Leibniz-Institut f\"ur Astrophysik Potsdam (AIP),  
Max-Planck-Institut f\"ur Astronomie (MPIA Heidelberg), 
Max-Planck-Institut f\"ur Astrophysik (MPA Garching), 
Max-Planck-Institut f\"ur Extraterrestrische Physik (MPE), 
National Astronomical Observatory of China, New Mexico State University, 
New York University, University of Notre Dame, 
Observat\'ario Nacional / MCTI, The Ohio State University, 
Pennsylvania State University, Shanghai Astronomical Observatory, 
United Kingdom Participation Group,
Universidad Nacional Aut\'onoma de M\'exico, University of Arizona, 
University of Colorado Boulder, University of Oxford, University of Portsmouth, 
University of Utah, University of Virginia, University of Washington, University of Wisconsin, 
Vanderbilt University, and Yale University.\\

This work has made use of data from the European Space Agency (ESA)
mission {\it Gaia} (\url{http://www.cosmos.esa.int/gaia}), processed by
the {\it Gaia} Data Processing and Analysis Consortium (DPAC,
\url{http://www.cosmos.esa.int/web/gaia/dpac/consortium}). Funding
for the DPAC has been provided by national institutions, in particular
the institutions participating in the {\it Gaia} Multilateral Agreement.
This work has also made use of data from {\it Gaia}-ESO based on data products from observations made with ESO Telescopes at the La Silla Paranal Observatory under programme ID 188.B-3002.\\

ABAQ and FA acknowledge support from the Leibniz Graduate School for Quantitative Spectroscopy at AIP, in particular the Outgoing and Incoming Mobility Programme. CC acknowledges support from DFG Grant CH1188/2-1 and from the ChETEC COST
Action (CA16117), supported by COST (European Cooperation in Science and Technology). TCB acknowledges partial support from grant PHY 14-30152: Physics Frontier Center/JINA Center for the Evolution of the Elements (JINA-CEE), awarded by the US National Science Foundation.



\bibliographystyle{mnras}
\bibliography{FA_library}


\appendix
\section {Additional simulation results}\label{appendixa}
\begin{figure*}
  \centering
  \includegraphics[width=15.5cm, trim=1cm 2.3cm 2cm 0cm, clip=true]{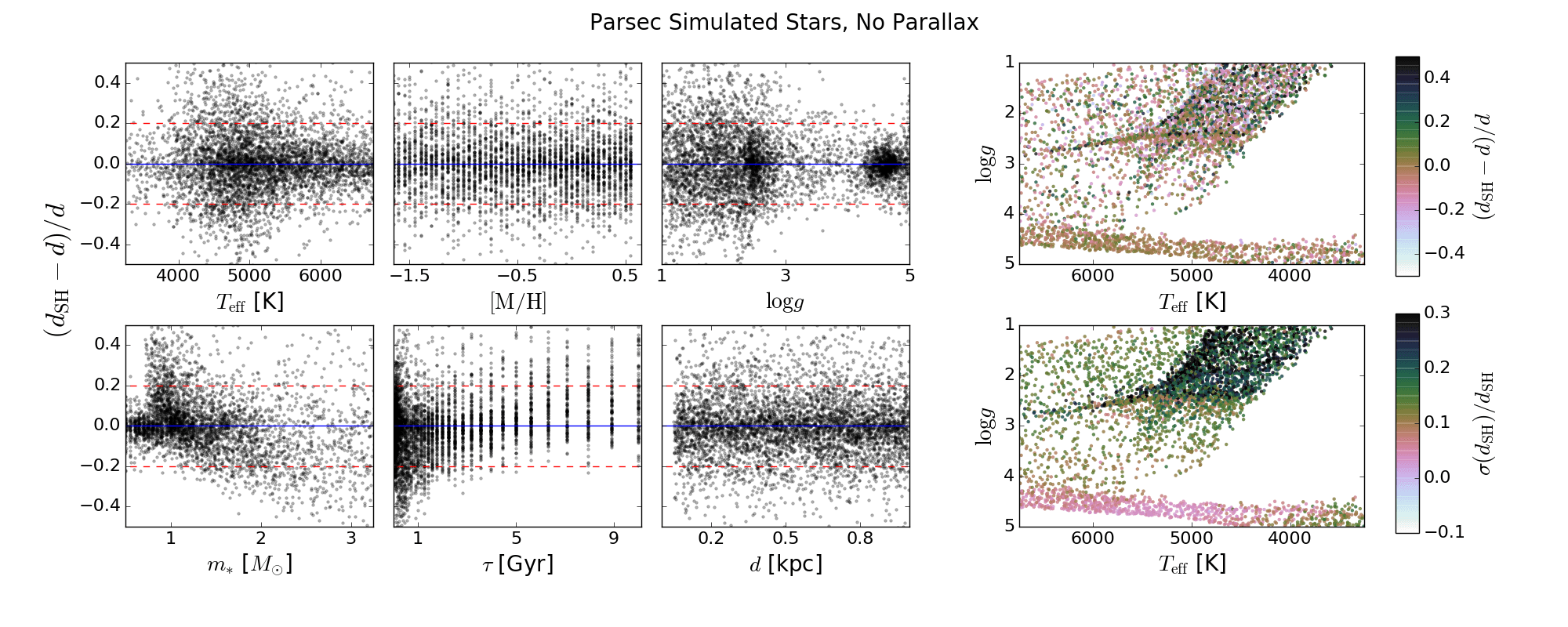}
  \includegraphics[width=15.5cm, trim=1cm 2.3cm 2cm 1.6cm, clip=true]{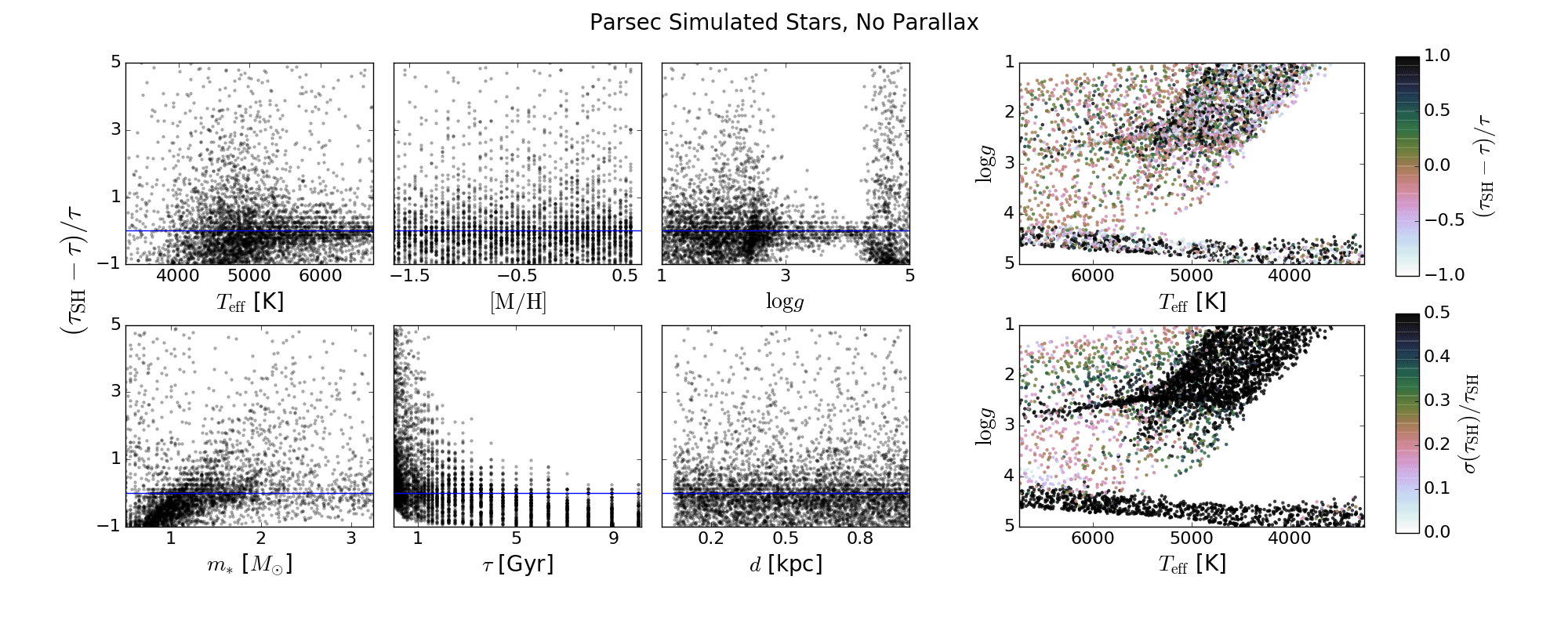}
  \includegraphics[width=15.5cm, trim=1cm 2.3cm 2cm 1.6cm, clip=true]{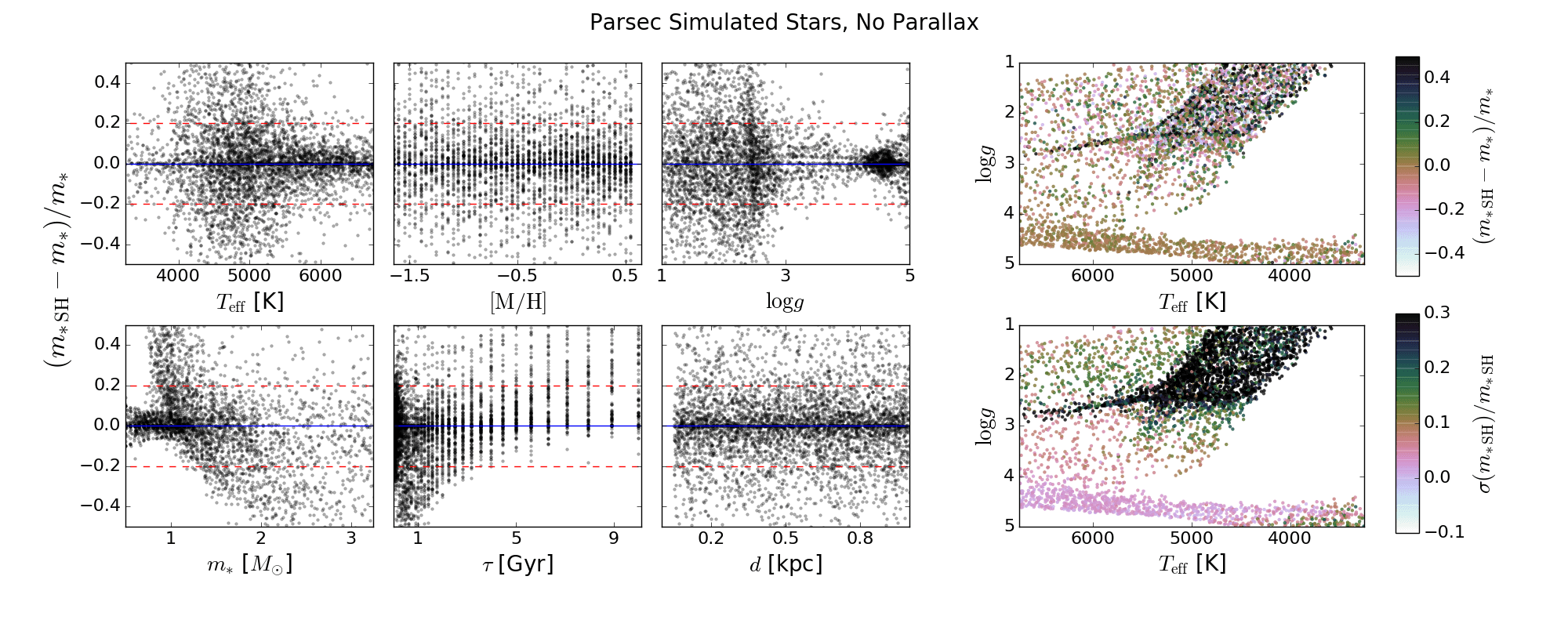}
  \includegraphics[width=15.5cm, trim=1cm 2.3cm 2cm 1.6cm, clip=true]{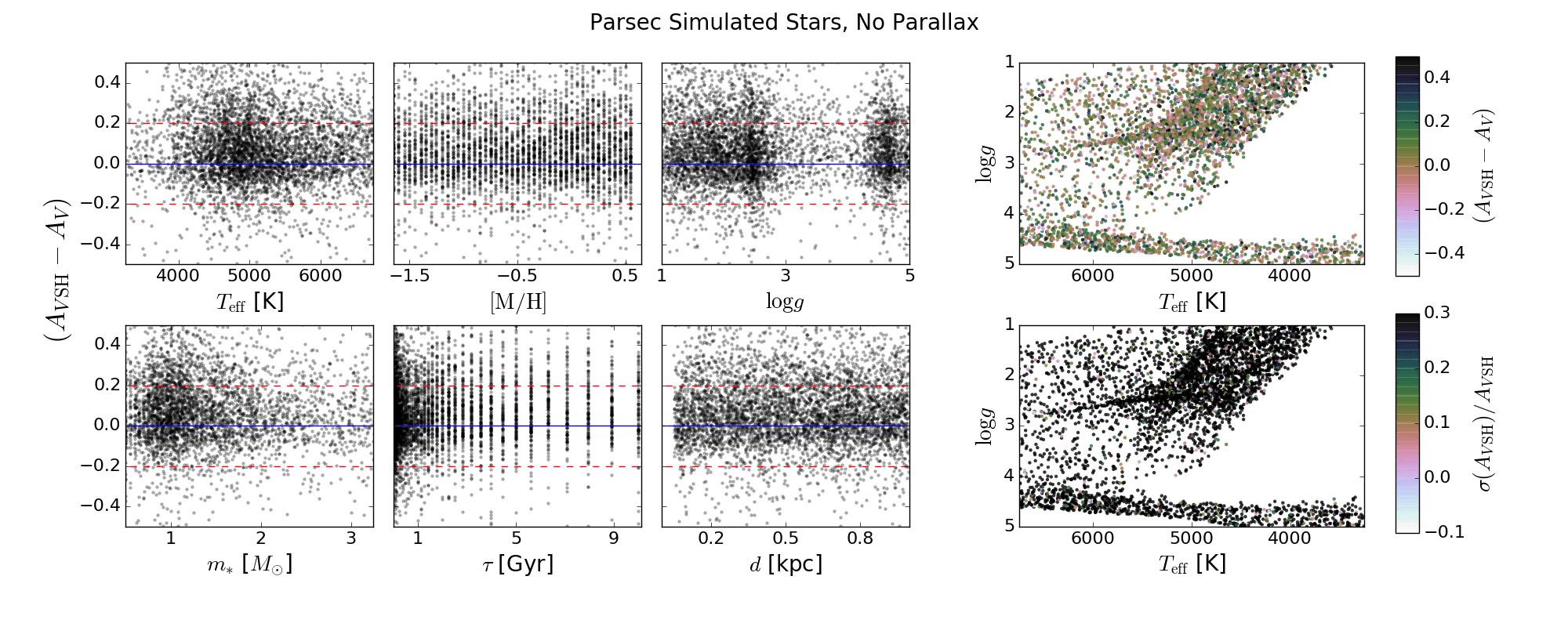}
  \caption{Same panels as in Figure \ref{parsec}, but now showing the results from {\tt StarHorse} when the constraint provided by the parallax is not used.} 
  \label{fakestarsnoparallax}
\end{figure*}

Here we show the results of PARSEC simulations when parallaxes are not used to constrain distances or the likelihood. Figure \ref{fakestarsnoparallax} can then be compared to the case shown in the main body of the paper, Figure \ref{parsec}. All estimated parameters are subject to larger errors, especially $A_V$. The systematic distance and mass error dependances on true stellar age and mass become very pronounced when parallaxes are not used.

\begin{figure*}
  \centering
  \includegraphics[width=15.5cm, trim=1cm 2.3cm 2cm 0cm, clip=true]{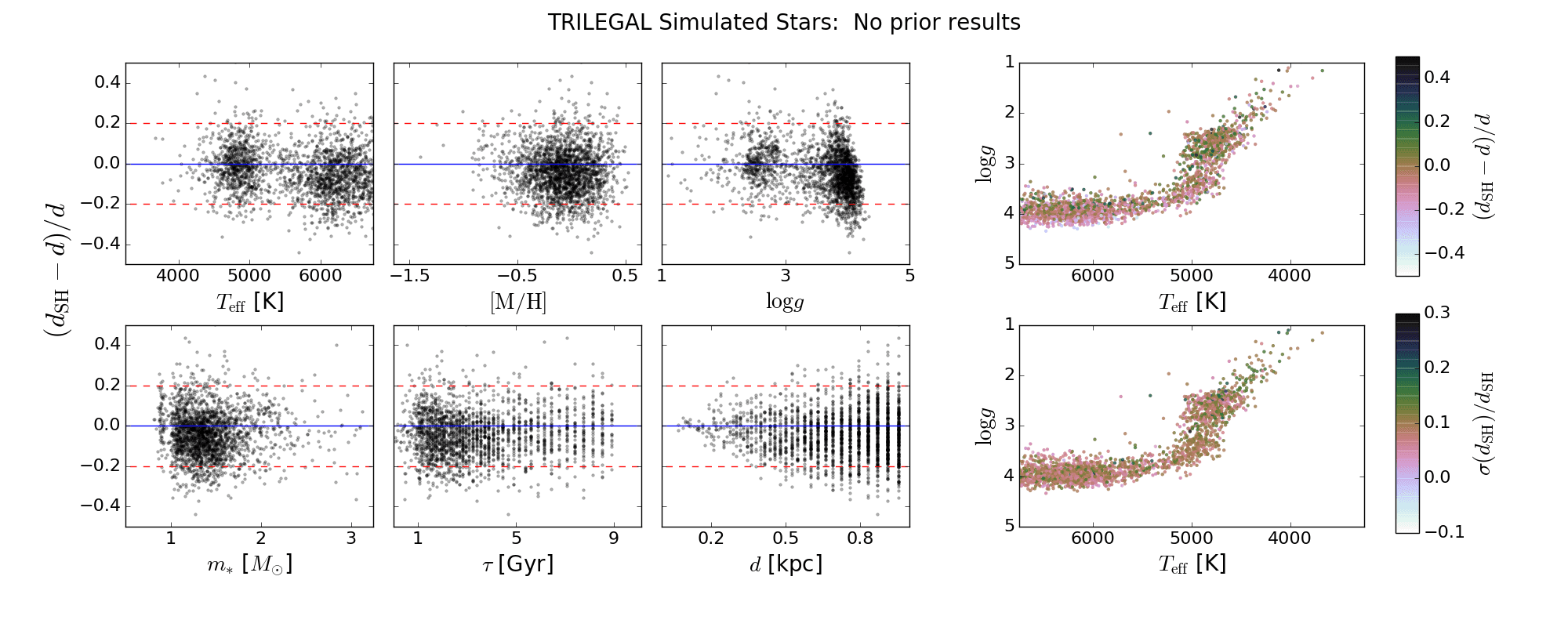}
  \includegraphics[width=15.5cm, trim=1cm 2.3cm 2cm 1.6cm, clip=true]{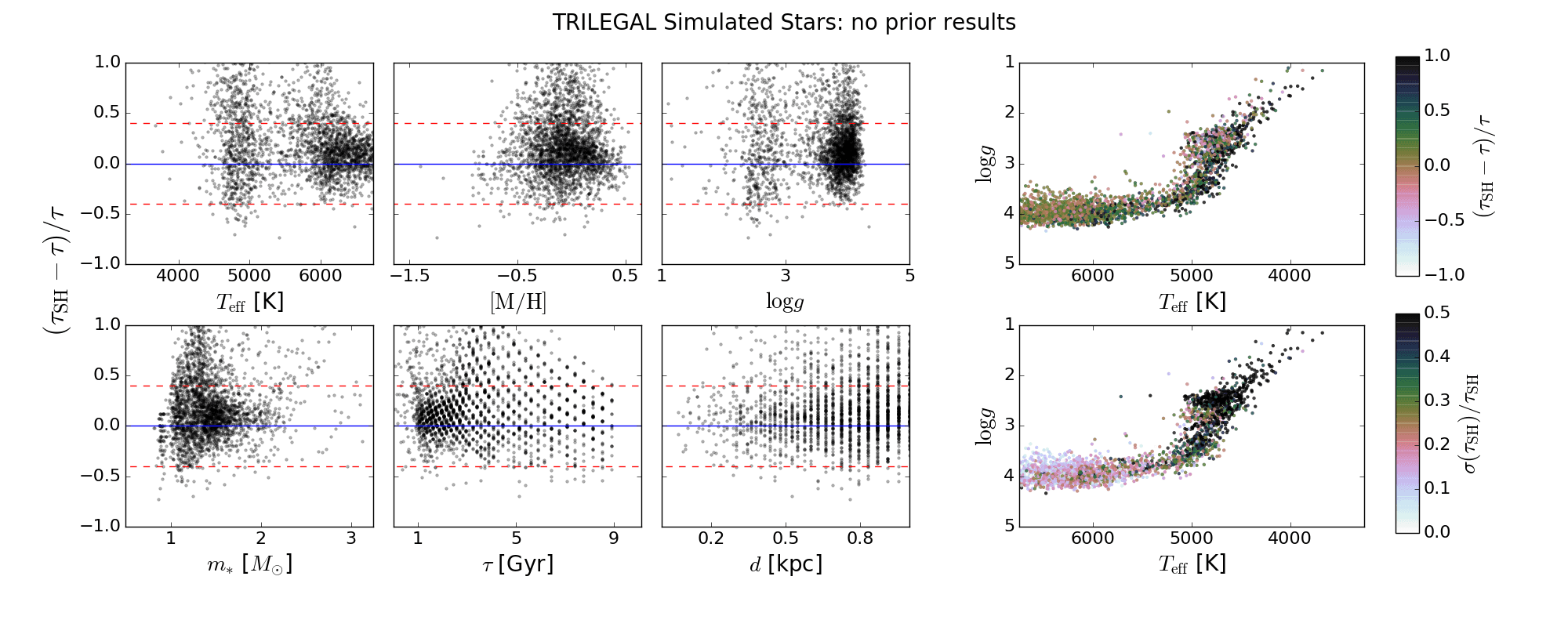}
  \includegraphics[width=15.5cm, trim=1cm 2.3cm 2cm 1.6cm, clip=true]{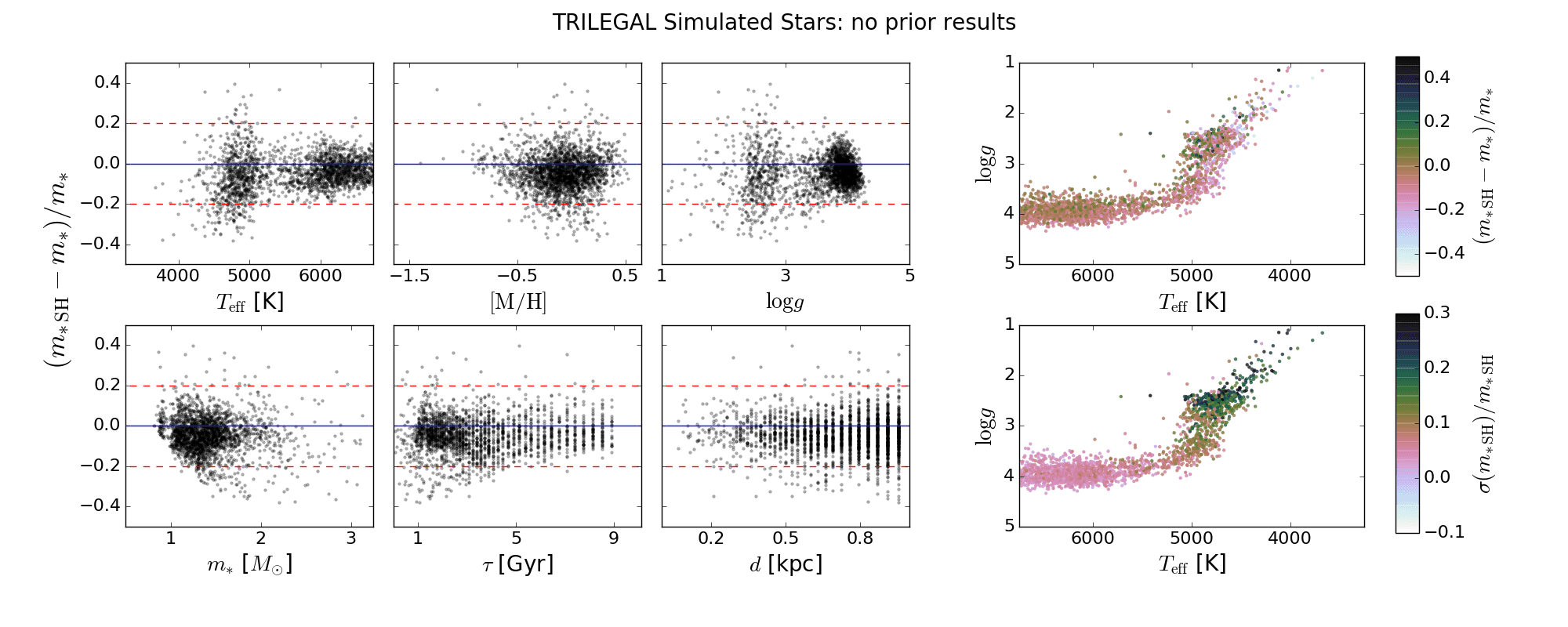}
  \includegraphics[width=15.5cm, trim=1cm 2.3cm 2cm 1.6cm, clip=true]{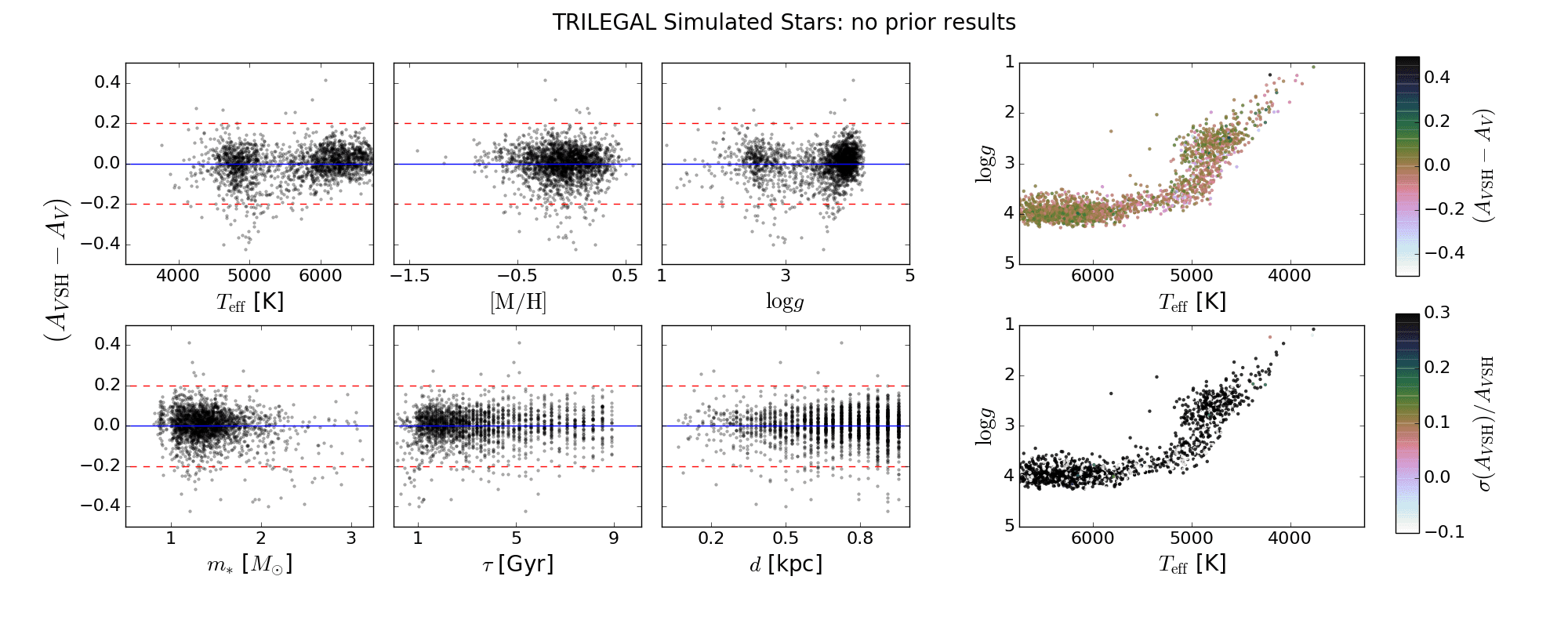}
  \caption{Same panels as in Figure \ref{trilegalfieldbulge}, but now showing the results from {\tt StarHorse} when no priors in metallicity, age and spatial distribution are adopted.} 
  \label{trilegalnoprior}
\end{figure*}

We also show the results of the TRILEGAL simulations for the case where no priors are adopted, to be compared to those shown in Figure \ref{trilegalfieldbulge}, for the {\it All priors} case. Figure \ref{trilegalnoprior} show the results for TRILEGAL simulations with no spatial, MDF or ADF see \S \ref{priors}. 

\section {Additional Data Released Analysis}

\begin{figure*}
\centering
  \includegraphics[width=.8\textwidth]{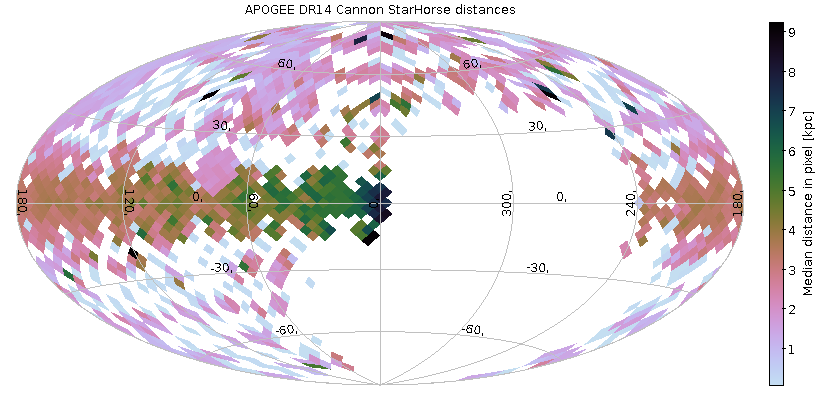} \\
  \includegraphics[width=.8\textwidth]{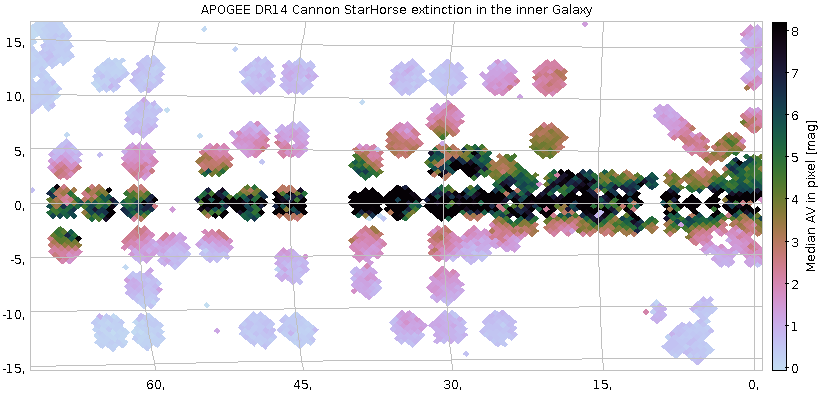} \\
  \includegraphics[width=.45\textwidth]{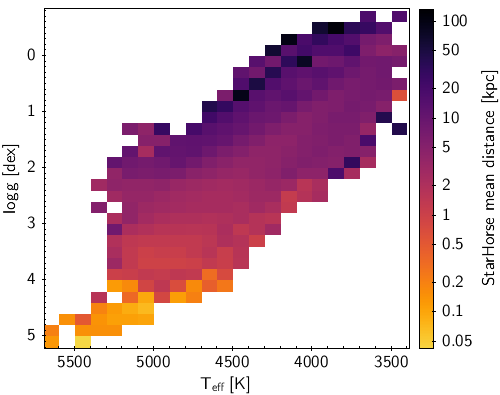} 
  \includegraphics[width=.45\textwidth]{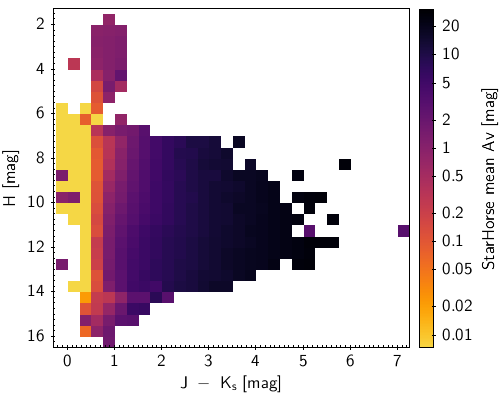} \\
  \caption{Illustration of the APOGEE DR14 Cannon distance and extinction results from {\tt StarHorse}. The panels and conventions are the same as in Figure 11.}
  \label{apocannon}
\end{figure*}

\begin{figure*}
\centering
  \includegraphics[width=.8\textwidth]{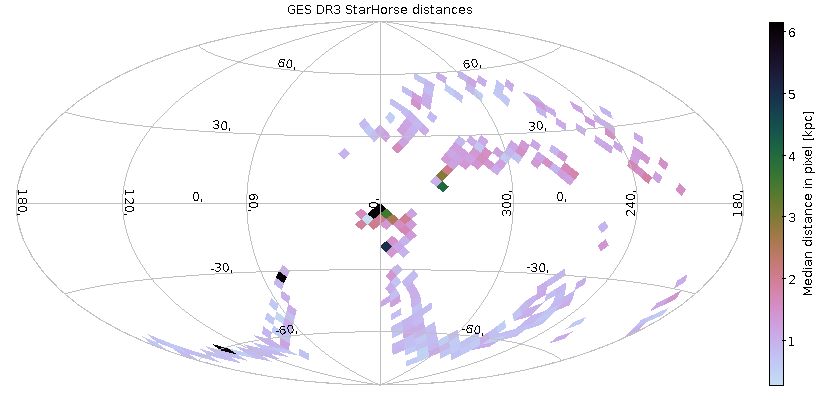} \\
  \includegraphics[width=.8\textwidth]{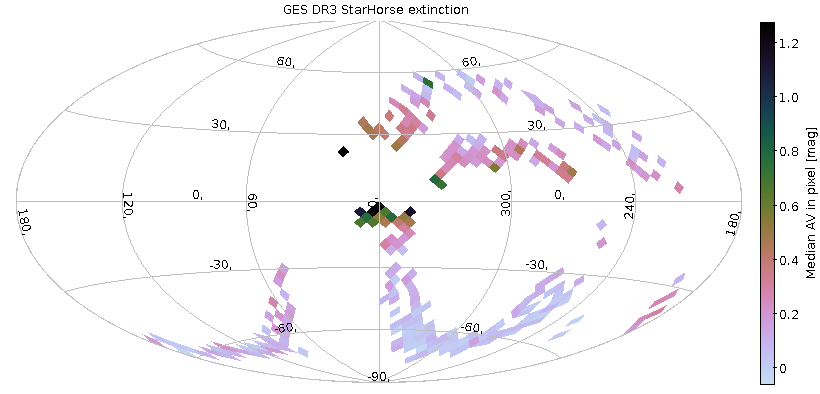} \\
  \includegraphics[width=.45\textwidth]{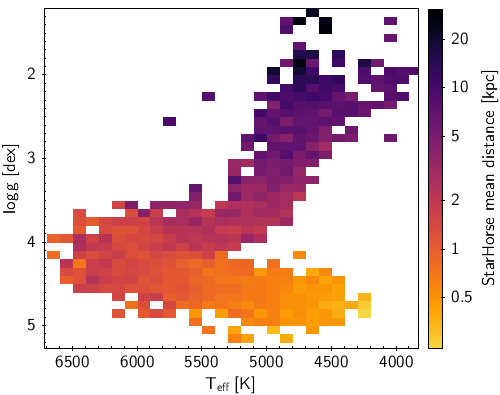} 
  \includegraphics[width=.45\textwidth]{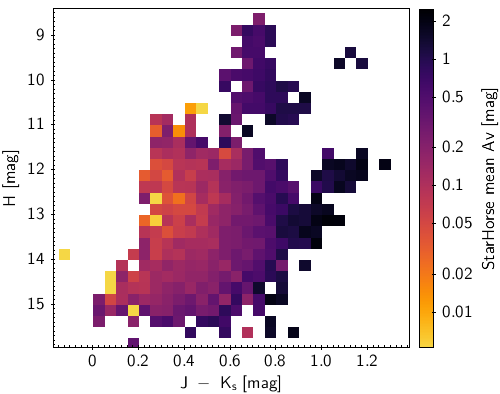} \\
  \caption{Illustration of the GES DR3 distance and extinction results from {\tt StarHorse}. The panels and conventions are the same as in Figure 11.}
  \label{fig:gesdr3}
\end{figure*}

\begin{figure*}
\centering
  \includegraphics[width=.8\textwidth]{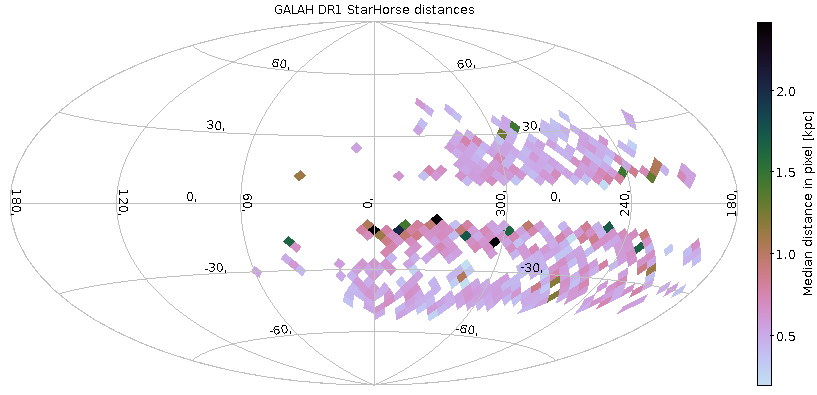} \\
  \includegraphics[width=.8\textwidth]{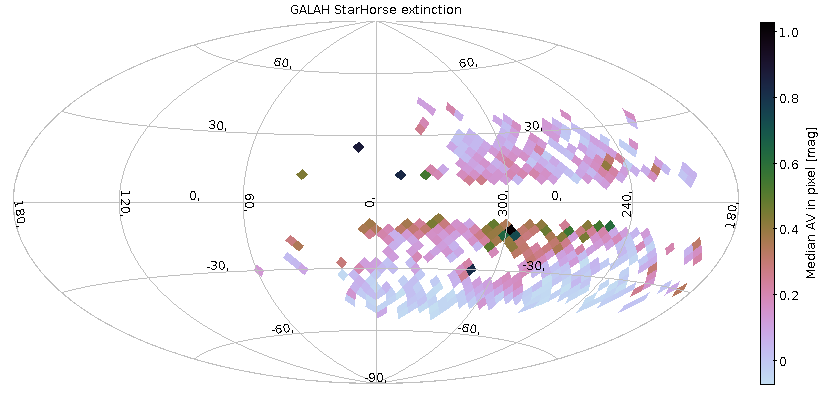} \\
  \includegraphics[width=.45\textwidth]{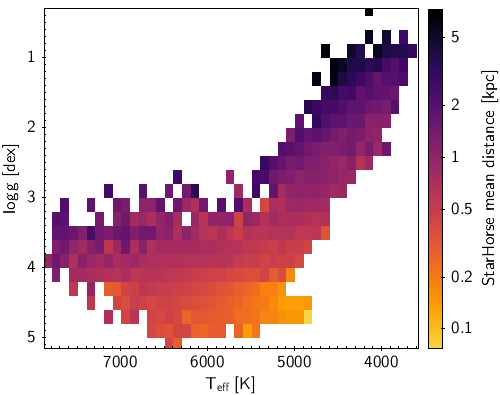} 
  \includegraphics[width=.45\textwidth]{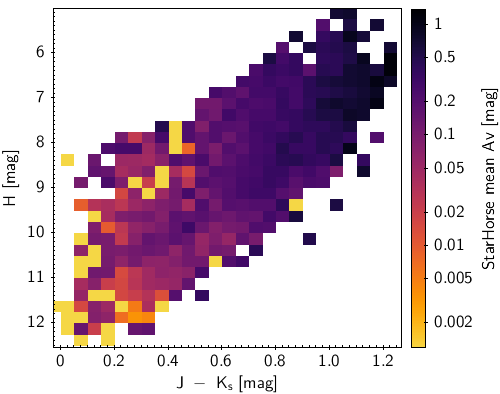} \\
  \caption{Illustration of the GALAH DR1 distance and extinction results from {\tt StarHorse}.The panels and conventions are the same as in Figure 11.}
  \label{fig:galahdr1}
\end{figure*}

\begin{figure*}
\centering
  \includegraphics[width=.8\textwidth]{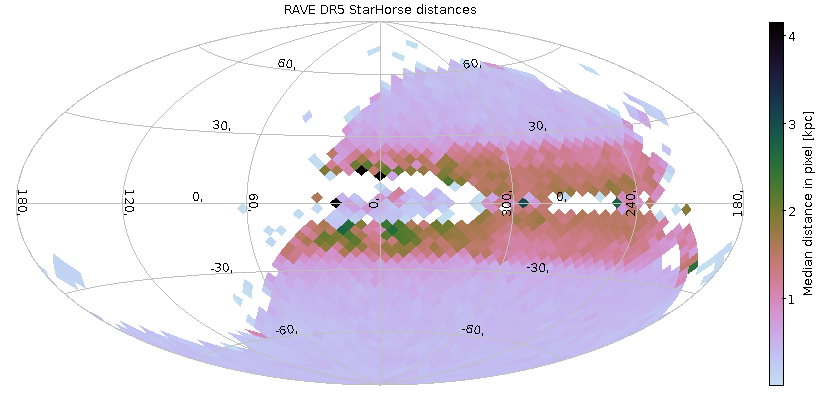} \\
  \includegraphics[width=.8\textwidth]{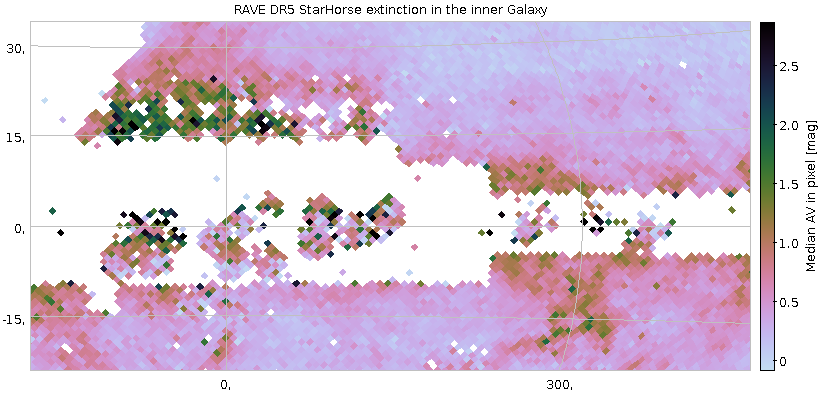} \\
  \includegraphics[width=.45\textwidth]{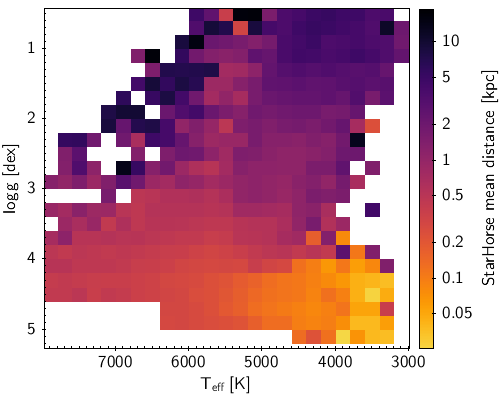} 
  \includegraphics[width=.45\textwidth]{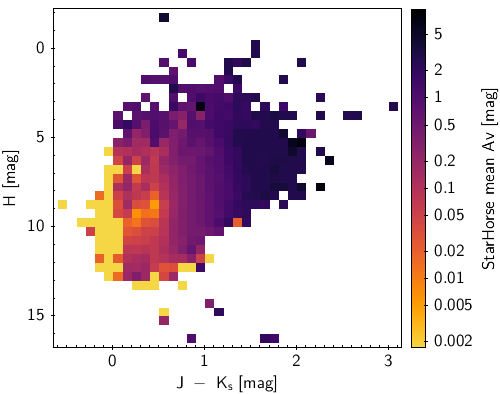} \\
  \caption{Illustration of the RAVE DR5 distance and extinction results from {\tt StarHorse}. The panels and conventions are the same as in Figure 11.}
  \label{fig:ravedr5}
\end{figure*}

\begin{figure*}
\centering
  \includegraphics[width=.8\textwidth]{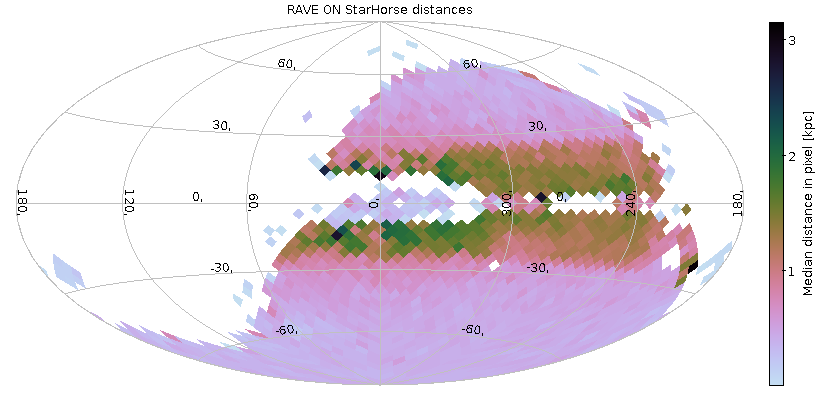} \\
  \includegraphics[width=.8\textwidth]{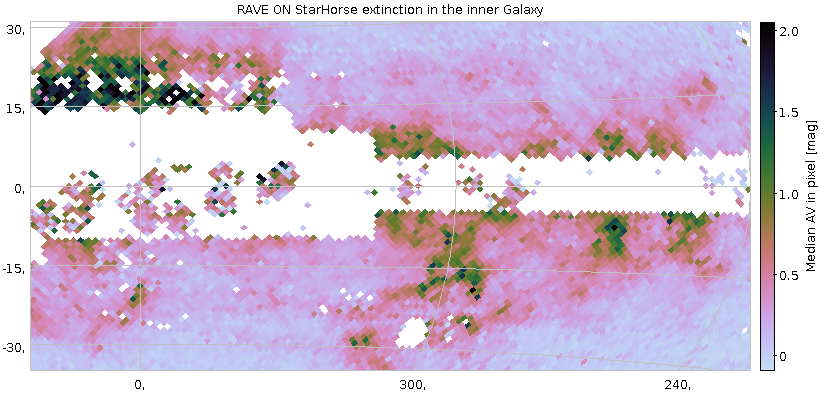} \\
  \includegraphics[width=.45\textwidth]{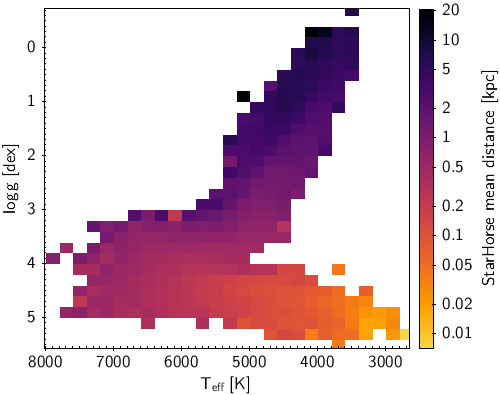} 
  \includegraphics[width=.45\textwidth]{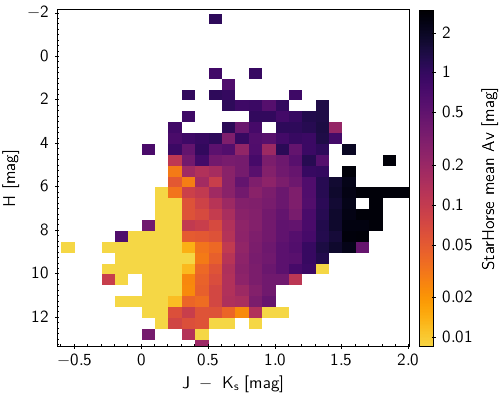} \\
  \caption{Illustration of the RAVE ON distance and extinction results from {\tt StarHorse}. The panels and conventions are the same as in Figure 11.}
  \label{fig:raveon}
\end{figure*}

\begin{figure*}
\centering
  \includegraphics[width=.8\textwidth]{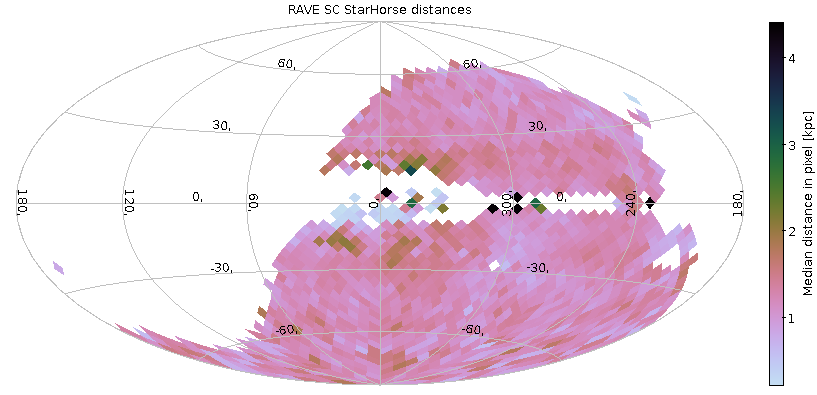} \\
  \includegraphics[width=.8\textwidth]{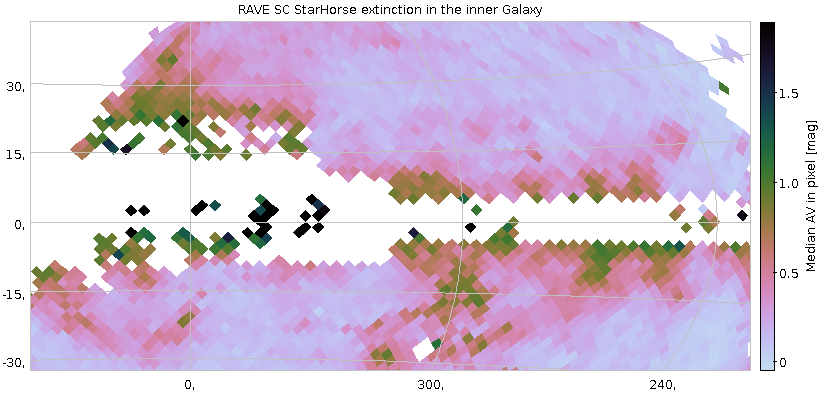} \\
  \includegraphics[width=.45\textwidth]{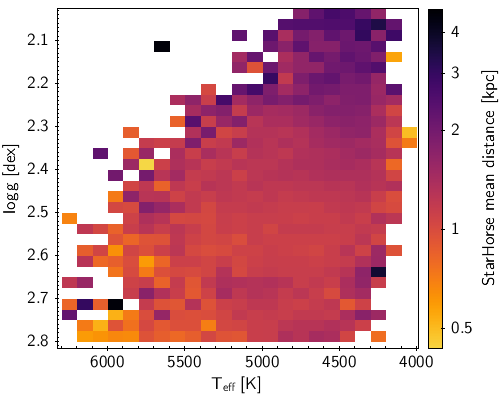} 
  \includegraphics[width=.45\textwidth]{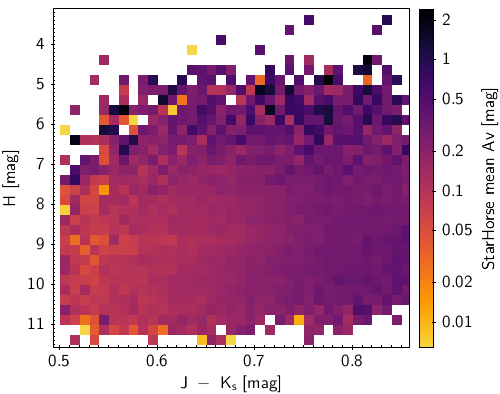} \\
  \caption{Illustration of the RAVE SC distance and extinction results from {\tt StarHorse}. The panels and conventions are the same as in Figure 11.}
  \label{fig:ravesc}
\end{figure*}


\bsp	
\label{lastpage}
\end{document}